\documentclass[12pt]{elsarticle}

\usepackage{fullpage}
\usepackage{verbatim}
\usepackage[usenames]{color}
\usepackage{subfigure}
\usepackage{epsfig}
\usepackage{amsfonts}
\usepackage{amssymb}
\usepackage{setspace}
\usepackage[ansinew]{inputenc}
\usepackage[cmex10]{amsmath}
\usepackage{psfrag}
\usepackage{wrapfig}
\usepackage[small]{caption}
\usepackage{graphicx}
\usepackage{lscape}
\usepackage{soul}

\newcommand{\Pmat}{\mathbf{P}}
\newcommand{\Phat}{\hat{\mathbf{P}}}
\newcommand{\Ptilde}{\tilde{\mathbf{P}}}
\newcommand{\pr}[1]{\Pr \left\{{#1}\right\}}

\def\piD{\boldsymbol{\pi}^{\text{d}}}
\def\piS{\boldsymbol{\pi}^{\text{s}}}
\def\pitildeS{\tilde{\boldsymbol{\pi}}_s}

\def\SIFS{\text{SIFS}}
\def\DIFS{\text{DIFS}}

\def\PER{\text{PER}}
\def\REF{\text{ref}}
\def\ACK{\text{ACK}}

\begin{document}
\newdimen\snellbaselineskip
\newdimen\snellskip
\snellskip=1.5ex
\snellbaselineskip=\baselineskip
\def\srule{\omit\kern.5em\vrule\kern-.5em}
\newbox\bigstrutbox
\setbox\bigstrutbox=\hbox{\vrule height14.5pt depth9.5pt width0pt}
\def\bigstrut{\relax\ifmmode\copy\bigstrutbox\else\unhcopy\bigstrutbox\fi}
\def\middlehrule#1#2{\noalign{\kern-\snellbaselineskip\kern\snellskip}
&\multispan#1\strut\hrulefill
&\omit\hbox to.5em{\hrulefill}\vrule
height \snellskip\kern-.5em&\multispan#2\hrulefill\cr}

\makeatletter
\def\bordermatrix#1{\begingroup \m@th
  \@tempdima 8.75\p@
  \setbox\z@\vbox{%
    \def\cr{\crcr\noalign{\kern2\p@\global\let\cr\endline}}%
    \ialign{$##$\hfil\kern2\p@\kern\@tempdima&\thinspace\hfil$##$\hfil
      &&\quad\hfil$##$\hfil\crcr
      \omit\strut\hfil\crcr\noalign{\kern-\snellbaselineskip}%
      #1\crcr\omit\strut\cr}}%
  \setbox\tw@\vbox{\unvcopy\z@\global\setbox\@ne\lastbox}%
  \setbox\tw@\hbox{\unhbox\@ne\unskip\global\setbox\@ne\lastbox}%
  \setbox\tw@\hbox{$\kern\wd\@ne\kern-\@tempdima\left(\kern-\wd\@ne
    \global\setbox\@ne\vbox{\box\@ne\kern2\p@}%
    \vcenter{\kern-\ht\@ne\unvbox\z@\kern-\snellbaselineskip}\,\right)$}%
  \null\;\vbox{\kern\ht\@ne\box\tw@}\endgroup}
\makeatletter

\makeatletter
\def\bordermatrix#1{\begingroup \m@th
  \@tempdima 8.75\p@
  \setbox\z@\vbox{%
    \def\cr{\crcr\noalign{\kern2\p@\global\let\cr\endline}}%
    \ialign{$##$\hfil\kern2\p@\kern\@tempdima&\thinspace\hfil$##$\hfil
      &&\quad\hfil$##$\hfil\crcr
      \omit\strut\hfil\crcr\noalign{\kern-\snellbaselineskip}%
      #1\crcr\omit\strut\cr}}%
  \setbox\tw@\vbox{\unvcopy\z@\global\setbox\@ne\lastbox}%
  \setbox\tw@\hbox{\unhbox\@ne\unskip\global\setbox\@ne\lastbox}%
  \setbox\tw@\hbox{$\kern\wd\@ne\kern-\@tempdima\left(\kern-\wd\@ne
    \global\setbox\@ne\vbox{\box\@ne\kern2\p@}%
    \vcenter{\kern-\ht\@ne\unvbox\z@\kern-\snellbaselineskip}\,\right)$}%
  \null\;\vbox{\kern\ht\@ne\box\tw@}\endgroup}
\makeatletter

\begin{frontmatter}

\title{Performance Analysis of CSMA/CA Protocols with Multi-packet Transmission}
\date{}

\journal{Computer Networks}

\author{B. Bellalta$^{(1)}$, A. Faridi$^{(1)}$, D. Staehle$^{(2)}$, J. Barcelo$^{(1)}$, A. Vinel$^{(3)}$, M. Oliver$^{(1)}$\\ $(1)$ Universitat Pompeu Fabra, Barcelona \\ $(2)$ University of Wuerzburg, Germany \\ $(3)$ Technology University of Tampere, Finland}

\begin{abstract}
Wireless objects equipped with multiple antennas are able to simultaneously transmit multiple packets by exploiting the channel's spatial dimensions. In this paper, we study the benefits of such Multiple Packet Transmission (MPT) approach, when it is used in combination with a Carrier Sense Multiple Access with Collision Avoidance (CSMA/CA) protocol for fully interconnected networks, addressing the interactions between the two mechanisms and showing the performance gains that can be achieved. To this end, a very simple Media Access Control (MAC) protocol that captures the fundamental properties and tradeoffs of a CSMA/CA channel access protocol supporting MPT is introduced. Using this protocol as a reference, a new analytical model is presented for the case of non-saturated traffic sources with finite buffer space. Simulation results show that the analytical model is able to accurately characterize the steady-state behavior of the reference protocol for different number of antennas and different traffic loads, providing a useful tool for understanding the performance gains achieved by MAC protocols supporting MPT.
\end{abstract}

\begin{keyword}
	CSMA/CA\sep Multi Packet Transmission\sep SDMA\sep Queueing Model\sep Non-saturation\sep Performance Evaluation
\end{keyword}

\end{frontmatter}

%
%

\section{Introduction} \label{sec:Intro}

MIMO techniques allow for a more efficient use of transmission resources by making the channel's spatial dimension available, using simple and effective signal processing techniques \cite{mietzner2009multiple}. This new spatial dimension can be used to enhance coverage and reliability (spatial diversity) or to boost the system throughput (spatial multiplexing) \cite{zheng2003diversity}. In this paper, we will focus on the spatial multiplexing feature, which can be used to transmit multiple packets at the same time, referred to as Multi-Packet Transmission (MPT). MPT by using Spatial Multiplexing can be seen as a packet-based extension of the Space Division Multiple Access (SDMA) or Multi-user MIMO concepts \cite{paulraj2003introduction}.

Although the benefits and drawbacks of MPT in point to point links are well-known, there is still a lack of results focusing on the new challenges and benefits that can arise by combining MPT capabilities with random-access MAC protocols. On the one hand, using MPT will reduce the number of transmission attempts and therefore decrease the collision probability in a contention scenario. On the other hand, in order to make the MPT work properly, the MAC protocol needs to be modified, at the expense of some extra temporal overhead, e.g., to include the necessary procedures for estimating the channel state between the transmitter and each receiver, required for isolating the different spatial streams, or to feed the CSI (Channel State Information) back to the transmitter when necessary. There will also be a need for extra acknowledgments to confirm the reception of all spatial streams. The negative impact of the extra temporal overhead imposed by these modifications on the system performance needs to be quantified and contrasted with the performance boost gained by enabling the simultaneous transmission of multiple packets.

This paper intends to provide insight on the interactions between CSMA/CA random access MAC protocols and the MPT scheme. We study and characterize these interactions by introducing a basic MAC protocol that combines these mechanisms and includes all the required features to make it work properly. The performance of the overall system is evaluated using an analytical model, whose accuracy is validated using simulations. The applied analysis strategy and the resulting model are general enough to provide deep insight into the analysis of any CSMA/CA MAC protocol with MPT capabilities.

In this context, the main contribution of this paper is to present a queueing model for wireless objects with MPT capabilities using a CSMA/CA protocol for channel access. We show that the considered approach provides a suitable path to modeling packet-based, multiple-antenna access protocols in non-saturation conditions. The presented results confirm the accuracy of the presented model and show the performance benefits of such protocols, as well as providing insights on how the system parameters have to be configured in order to achieve better performance gains.

CSMA/CA-based MAC protocols with MPT capability have been recently considered for WLANs due to their ability to improve the network performance, while keeping both the simplicity and the efficiency that random access MAC protocols possess \cite{li2010multi,cha2012performance}. In addition, the upcoming appearance of the IEEE 802.11ac amendment for WLANs \cite{IEEE80211ac}, which only enhances the Access Points with the MPT capability, is also pushing in that direction.

In \cite{li2010multi}, a MAC protocol for WLANs supporting MPT at the Access Point (AP) is presented. The RTS/CTS handshake procedure is extended to both coordinate the transmission of multiple packets from the AP to different STAs, and to provide the AP with the required CSI. An analytical model of the presented protocol in non-saturation conditions is introduced, although the authors only focus on the MAC performance, as queueing is not included in the model, hence not allowing for metrics such as average waiting delay per packet or the packet-loss probability due to buffer overflow to be computed.  In \cite{cha2012performance}, the performance of MPT in the downlink for the upcoming IEEE 802.11ac standard is evaluated. An accurate physical-layer and channel model is provided when a zero-forcing precoding scheme \cite{paulraj2003introduction} is used to create the multiple spatial streams. To evaluate the system performance, an analytical model in saturation conditions is presented, where it is assumed that the AP can always transmit as many packets as the number of antennas it has, thus also not considering the queueing dynamics.

However, MPT can be also applied to a wider set of scenarios in which the use of CSMA/CA MAC protocols is common, such as in Wireless Multimedia Sensor Networks that may have high bandwidth demands to transmit audio and video signals \cite{akyildiz2007survey}, in cluster-based Wireless Networks \cite{azadeh2011performance}, to enhance the cluster-head capabilities to transmit to multiple nodes simultaneously, or in vehicular networks \cite{mecklenbrauker2011vehicular}, to improve the car-to-car communication, among others. As these scenarios usually include a large number of nodes contending for the channel, they are more interesting from the point of view of the interactions between the MPT and CSMA/CA protocols, and for that reason we focus on them in this paper.

The rest of the paper is structured as follows. Section \ref{sec:Scenario} presents the considered reference scenario, i.e., a single-hop network where $N$ nodes compete for the channel. Section \ref{sec:MACMA} details the reference protocol. In Section \ref{sec:model}, the analytical model is presented. The analytical model validation and performance results are discussed in Section \ref{sec:Results}. Finally, the main conclusions of the paper are summarized in Section \ref{sec:Conclusions}.

%
%

\section{System Model} \label{sec:Scenario}

\begin{figure}[t!!!!!!!!]
\centering
\epsfig{file=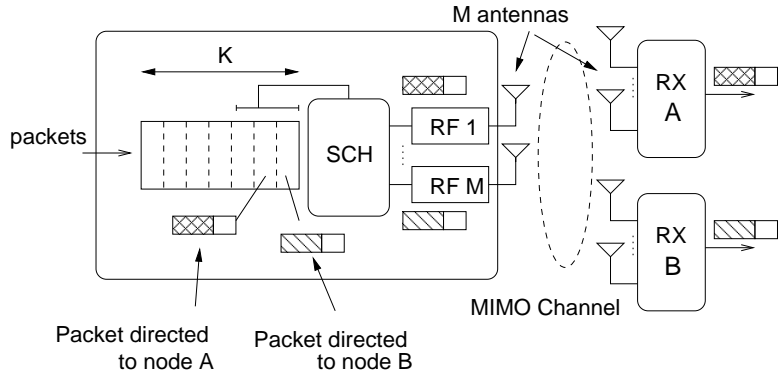,scale=0.6,angle=0}
\caption{A node transmitting two packets simultaneously {to two different destinations}}
\label{Fig:NodeScheme}
\end{figure}

A single-hop network with $N$ nodes is considered. Each node is assumed to be within the transmission range of all the other nodes and close enough to have negligible propagation delay. Every node is equipped with $M$ antennas as shown in Figure \ref{Fig:NodeScheme}, which allow it to simultaneously transmit up to $M$ packets to a single or multiple destination nodes.

\subsection{Node Operation and Link Layer}

Each node has a finite buffer of length $K$ packets, to which packets arrive according to a Poisson process of rate $\lambda$. Each packet has a fixed length of $L$ bits and can be directed to any of the other nodes. Packets depart from the node in batches, called space-batches, and are selected for transmission following a First-In First-Out (FIFO) policy, regardless of their destination(s). The space-batches are scheduled immediately following departure instants, i.e., when the previous space-batch is purged from the queue. The number of packets included in a space-batch, $s(q)$, depends on $q\in[0,K]$, the queue occupancy at the moment the new transmission is scheduled, and two system parameters $s_{\min}$ and $s_{\max}$, as follows:
\begin{equation}\label{Eq:batch_policy}
    s(q) = \left \{\begin{array}{lr}
			s_{\min}, &	q < s_{\min} \\
			q, 		&	s_{\min} \leq q < s_{\max}\\
			s_{\max},	& 	s_{\max} \leq q \\
          \end{array}\right.
\end{equation}
which can be written more concisely as $s(q)=\max\{s_{\min},\min\{q,s_{\max}\}\}$. Note that if just after a departure $q<s_{\min}$, the scheduler waits until enough packets have arrived and a space-batch containing $s_{\min}$ packets can be constructed. The parameters $s_{\min}$ and $s_{\max}$ can take values between $1$ and $M$, with $s_{\min}\leq s_{\max}$. They are design parameters that can be carefully chosen, depending on the arrival rate and the channel conditions, to improve the system performance. Choosing a high value for $s_{\min}$ reduces the number of transmission attempts on the channel, thus reducing the probability that a transmission results in a collision but at the cost of a larger average waiting delay, specially at low traffic rates. The $s_{\max}$ value has to be adjusted considering the channel state. In general, for any multi-user beamforming scheme, under good channel conditions (high Signal-to-Noise Ratio (SNR)), $s_{\max}$ can be increased towards $M$ as the system can benefit from a larger number of parallel transmissions with reasonably low transmission error probability. However, at low SNR, high $s_{\max}$ values may result in a high packet error rate (\PER), hence limiting the system throughput.

The channel access is governed by the reference MAC protocol detailed in Section \ref{sec:MACMA}. It is based on the IEEE 802.11 Distributed Coordination Function (DCF) \cite{IEEE80211b}. Basically, it differs from the IEEE 802.11 DCF in the following aspects: $1$) it relies on a single-stage backoff mechanism, $2$) a training sequence for each antenna needs to be transmitted in order to assess the CSI corresponding to that antenna, $3$) multiple packets can be transmitted in parallel, each using a different antenna, and $4$) an ARQ protocol is implemented to handle packet acknowledgements and retransmissions when multiple packets are transmitted in parallel.

\subsection{Operation of Other Layers}

Since the focus of this paper is on the link layer (transmission queue and MAC protocol) and in particular on an analytical model for computing the impact of link layer parameters on the system performance, a simple channel model, physical layer, and source traffic model are assumed in the system model.

\subsubsection{Channel Model}

A quasi-stationary channel with flat fading coefficients changing independently from transmission to transmission (MIMO Rayleigh channel) is considered. In addition, regardless of the specific positions of the transmitting and receiving nodes, it is assumed that for every transmitted packet, all the receivers observe the same average SNR.

\subsubsection{Physical Layer}

For transmitting a space-batch of size $m$, the physical layer, after receiving the $m$ selected packets from the link layer, builds the space-batch frame by mapping each scheduled packet to an antenna and encoding and modulating the signal for its transmission over the channel.
Equal transmission power of $P_{\text{t}}/m$ is allocated to each packet, where $P_{\text{t}}$ is the total transmission power. Two transmission rates are considered; one for the physical layer headers, $r_{\text{phy}}$, and the other for the MAC protocol headers and data, $r$.

In reception, the physical layer estimates the channels and detects the different spatial streams using a Zero-Forcing (ZF) detector \cite{paulraj2003introduction}, which are successively demodulated and decoded. The average received SNR at each node is $\xi_0$, and $\xi_0/m$ is the average received SNR at any receive antenna contributed by a single transmit antenna.

Considering the previous settings and a ZF detector, the post-processing SNR at each spatial stream is usually assumed independent of the other branches and follows a $\chi^2$ distribution with $l=2\cdot (M - m + 1)$ degrees of freedom, where $M$ is the number of receiving antennas and $m$ the number of simultaneously transmitted spatial streams \cite{paulraj2003introduction}. Based on this, and assuming that a packet is received correctly only if the post-processing SNR is higher than a certain reference SNR, $\xi_{\REF}$, the Packet Error Rate (PER) when the transmission contains $m$ spatial streams is equivalent to the probability that the post-processing SNR, $\xi$, is smaller than $\xi_{\REF}$, i.e.,

\begin{equation}\label{Eq:PER_SINR}
   \PER(m)=F_{\xi}(\xi_{\REF})=1-\sum_{k=0}^{M-m}{\frac{1}{k!}\left( \frac{m}{\xi_0}\xi_{\REF} \right)^{k}e^{-\frac{m}{\xi_0}\xi_{\REF}}}
\end{equation}
where $F_{\xi}$ is the cumulative distribution of the post-processing SNR. Note that, given a fixed number of antennas, the \PER~increases with the number of spatial streams transmitted, thus showing the existence of a reliability-capacity tradeoff \cite{zheng2003diversity}. The capture effect is not considered and in case of collision, all transmitted packets are assumed to be corrupted and cannot be decoded correctly.

%
%
%

\section{The Reference MAC Protocol} \label{sec:MACMA}

The reference MAC protocol shares with the DCF the channel access and backoff techniques, and the meaning and functionality of parameters such as the DIFS (Distributed Inter-frame Space) and SIFS (Short Inter-frame Space).

\subsection{Protocol Description}

At any given time, a node operating under the reference protocol can be in either of the following operation modes: \textit{idle}, \textit{transmission}, and \textit{reception}.

\subsubsection{Idle Mode}

A node is in idle mode if it is neither transmitting nor receiving. From the idle mode, the protocol will move to the reception mode if the channel is detected busy, or to the transmission mode if the number of packets stored in its queue reaches $s_{\min}$ due to new arrivals.

\subsubsection{Transmission Mode}

As soon as a node enters transmission mode, it builds a new space-batch using the first $s(q)=\max\{s_{\min},\min\{q,s_{\max}\}\}$ packets stored in its queue. It will then wait until the channel is detected free for a period equal to a DIFS, after which it initiates a random backoff to avoid collision with other nodes' transmissions. When the backoff ends, the space-batch is transmitted and the transmitting node waits for the corresponding ACKs.

The backoff value is selected according to a uniform distribution in the range $[0,\text{CW}-1]$, with $\text{CW}$ the contention window length. If the channel is detected free, the backoff counter is decreased by one unit until it reaches $0$, at which instant the space-batch is transmitted. Selecting a random value equal to $0$ means immediate transmission. If the channel is detected busy at the beginning of a backoff slot, the countdown is frozen until the channel becomes idle again.

A space-batch transmission can have three different outcomes: it can be successful, end in collision, or arrive collision-free but contain erroneous packets. We say that a packet in the space-batch is transmitted successfully if its reception is acknowledged by the receiver, which means that it has neither suffered a collision nor contains errors.

The packets corresponding to a space-batch will be purged from the queue only when the reception of all packets in that space-batch have been acknowledged. When a packet is not correctly received, due to either errors or collisions, in the next transmission, only the remaining (unacknowledged) packets of the previous space-batch will be transmitted. This retransmission procedure is reiterated until all the packets of the original space-batch are correctly received. In other words, the newly admitted packets to the queue will not be transmitted within the retransmitted space-batches, even if the number of retransmitted packets included in the space-batch is lower than $s_{\max}$. The motivation of this approach is that, the fewer the number of packets transmitted in a space-batch, the higher the chances that they do not suffer transmission errors and therefore the expected number of retransmissions due to errors is kept low.

\begin{figure*}[t!!!!!!!!]
\centering
\epsfig{file=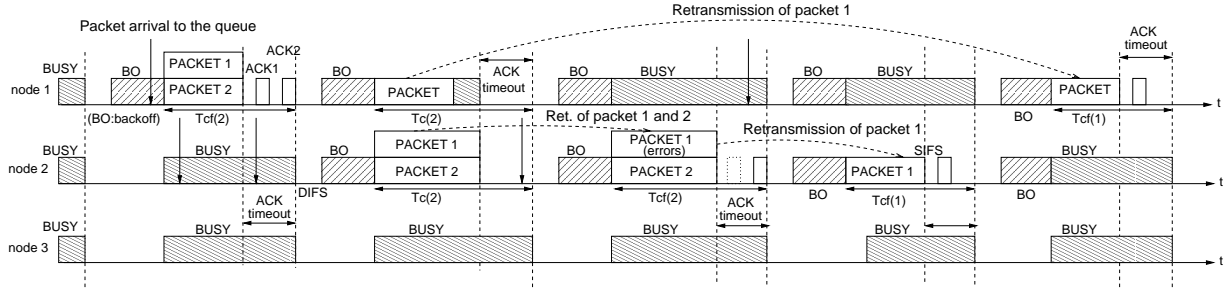,scale=0.35,angle=0}
\caption{The reference protocol operation with $s_{\min}=1$, $s_{\max}=2$ and $M=2$: successful transmission and collision duration}
 \label{Fig:MAC_operation_col}
\end{figure*}

Figure \ref{Fig:MAC_operation_col} shows an example of the operation of the reference protocol for both cases of a collision-free transmission and a collision. As shown in the figure, a separate ACK is transmitted for every successfully received packet. The considered retransmission mechanism can be observed in the third transmission of the second node in Figure \ref{Fig:MAC_operation_col}, where only the unacknowledged packet from the previous transmission is sent, although there are more packets waiting in the queue.

The time during which the channel remains busy due to a collision-free transmission a space-batch of size $m$, $T_{\text{cf}}(m)$, regardless of whether or not it contains errors, is given by
\begin{equation}\label{Eq:tx_succ_duration}
  T_{\text{cf}}(m)=T_{\text{data}}(m)+\ACK_{\text{timeout}}
\end{equation}
where $T_{\text{data}}(m)$ is the duration of the data portion of the transmitted (or retransmitted) frame that includes $m$ packets, and $\ACK_{\text{timeout}}$ is the time spent waiting for and receiving the ACK for each successfully transmitted packet. Note that, to avoid confusion, we use the notation $s(q)$, or simply $s$, to refer to the original space-batch size, and $m \leq s$, to refer in general to the number of packets involved in a transmission.

It is important to note here that $\ACK_{\text{timeout}}$ has to be set to its maximum possible value, i.e.,
\begin{equation}\label{Eq:ACK_timer}
  \ACK_{\text{timeout}}=M \cdot \left(\SIFS+T_{\text{ack}}\right)
\end{equation}
where SIFS is the Short Inter-Frame Space and $T_{\text{ack}}$ is the duration of an ACK frame. This is because there can always be nodes that are not able to decode the space-batch due to errors and therefore, cannot know the number of packets that it includes. Furthermore, a node that does receive the space-batch successfully has no way of knowing whether all other nodes have been able to do the same. Since it is required that all nodes restart the channel contention at the same time after a transmission, all nodes will have to wait the total time required for receiving $M$ acknowledgements.

Similarly, in case of a collision, both the transmitter and the nodes that have overheard the transmission have to wait for a full $\ACK_{\text{timeout}}$ in order to receive all possible ACKs. Additionally, as space-batches of different durations, carrying a different number of packets, can collide, the $\ACK_{\text{timeout}}$ has to be initiated only after the transmission of the data part of the longest space-batch involved in the collision. This guarantees that all nodes will restart to contend for the channel at the same time. Therefore, the collision duration is given by
\begin{equation}\label{Eq:tx_col_duration}
  T_{\text{c}}(m)=T_{\text{data}}(m_{\max})+ \ACK_{\text{timeout}}
\end{equation}
where $m_{\max}$ represents the longest space-batch involved in the collision. 

Note that, according to the reference protocol, the transmitter does not have access to the CSI, and therefore, it is not able to decode the ACKs if they are simultaneously sent from all the receivers. If this was possible, the $\ACK_{\text{timeout}}$ could be significantly shorter, resulting in an improvement in the overall system performance. However, this would be at the cost of having extra overheads necessary to feed the transmitter with the CSI from each receiver, e.g., by using a modified RTS/CTS mechanism \cite{li2010multi}.

\subsubsection{Reception Mode}

A node in reception mode checks the receiver address of each correctly received packet from the arriving space-batch frame. For each packet that is directed to it, it schedules the transmission of an ACK following the order indicated in the headers of each data packet. This means that the $i$-th ACK is transmitted $\SIFS+(i-1)\left(T_{\text{ack}}+\SIFS\right)$ seconds after the reception of the space-batch. After the $\ACK_{\text{timeout}}$, all nodes move to the idle or remain in the transmission mode, depending on their queue status.

\subsection{Duration of a Space-batch transmission} \label{sec:frame_structure}

The duration of the space-batch frame is variable and depends on the number of training
sequences included. Therefore, the duration of a space-batch frame is given by:
\begin{equation}
\begin{small}
    T_{\text{data}}(m)=\left \{\begin{array}{lr}
           \frac{L_{\text{phy}}}{r_{\text{phy}}}+\frac{L_{\text{mac}}+L}{r}, & m=1 \\
    	   \frac{L_{\text{phy}}+L_{\text{tr}}(m)}{r_{\text{phy}}}+\frac{L_{\text{mac}}+L}{r}, &
m\in[2,M] \\	
          \end{array}\right.
\end{small}
\end{equation}
where $L_{\text{phy}}$, $L_{\text{tr}}(m)$, and $L_{\text{mac}}$ are the PHY header, the variable
number of training sequences, and the MAC header length, respectively.

A space-batch transmission starts with the PHY header. After it, $m$ training sequences are transmitted sequentially, allowing the receiver to estimate the fading coefficients between the transmitting antennas and all its receiving antennas, which are used to separate the received packets. For the specific case of $m=1$, no training sequence is required for the receiver, as there is no need to detect multiple transmitted streams. Then, the MAC header and the data packets are transmitted.

The ACK has a fixed duration equal to

\begin{equation}
\begin{small}
T_{\text{ack}}=\frac{L_{\text{phy}}}{r_{\text{phy}}}+\frac{L_{\text{mac}}}{r}
\end{small}
\end{equation}

%
%

\section{Analytical Modeling of the Protocol} \label{sec:model}

{To analyze the behavior of a node in a network operating under the reference protocol, a queueing model is built that captures the random packet arrivals at the node, its MPT capability, the protocol operation, and the presence of a finite-space buffer. A Poisson process is considered for the packet arrivals, and the departure process is analyzed by taking into account the interactions among nodes when accessing the shared channel using the reference protocol.} The model is then solved numerically by applying a fixed-point approach. Given an initial set of parameters, the fixed-point algorithm iterates until the convergence condition is reached (i.e., all the parameters of the model stabilize to a specific value).

In this work, in order to analytically model the MPT capability in non-saturation conditions, we use a batch-service queueing model \cite{gross1998fundamentals} for analyzing the queueing dynamics. Batch-service queuing models are currently receiving a lot of attention due to their application in modeling various features of communication systems, such as packet aggregation schemes on a temporal scale \cite{Bellalta2009-ASMTA}. Examples of use in specific technologies are \cite{kuppa2006modeling,lu2007performance,lu2009performance} focusing on the performance analysis of temporal aggregation in Wireless Local Area Networks (WLANs).

\subsection{MPT Queueing Model}

A batch-service $M/G^{[\boldsymbol{s}]}/1/K$ queue \cite{gross1998fundamentals} is used to model each node. In this notation, $\boldsymbol{[s]}=[s_{\min},s_{\max}]$ indicates an interval, with $s_{\min}$ and $s_{\max}$ representing the minimum and maximum number of spatial streams that can be used at each transmission, respectively. Poisson arrivals of rate $\lambda$ and a general service time distribution are considered. The buffer has a size of $K$ packets and no extra space is considered for the packets in service, and therefore, the packets included in a space-batch transmission remain stored in the queue until they are correctly received and acknowledged.

As mentioned before, packets are served in batches of size \mbox{$s(q) = \max\{s_{\min},\min\{q,s_{\max}\}\}$}, where $q$ is the number of packets present in the queue immediately after a departure. Here a {\it departure} refers to the moment at which all packets from the previous space-batch are acknowledged and purged from the queue. The scheduling of a space-batch takes place after each departure, as soon as enough packets become available in the queue.

Let $q_n$ denote the queue occupancy at the $n$-th departure instant. Then $q_n$ evolves according to the following recursion:
\begin{equation}\label{Eq:queue_recursion}
    q_{n} = \min \left\{V[s(q_{n-1})]+[q_{n-1} - s_{\max}]^+ ,K-s(q_{n-1})\right\}
\end{equation}
where the notation $[x]^+$ represents the non-negative part of $x$, and $V[s(q_{n-1})]$ is the number of packet arrivals during the transmission
phase of the $(n-1)$-st inter-departure epoch, i.e., during the time required to transmit $s(q_{n-1})$ packets, from the moment they are scheduled until they are purged from the queue. Therefore, $V[s(q_{n-1})]+[q_{n-1} - s_{\max}]^+$ is the number of packets that would be present in the queue at the end of a departure if the buffer had infinite capacity, and $K-s(q_{n-1})$ is the maximum possible queue occupancy immediately after a departure.

In order to find analytical formulation for key performance metrics, such as
delay and throughput, the steady-state queue occupancy probabilities need to be calculated. To derive these probabilities, first, the steady-state distribution for the queue states immediately after departures, $\piD$, is derived using a discrete-time embedded Markov chain. Then the PASTA (Poisson Arrivals See Time Averages) property of the Poisson arrivals \cite{gross1998fundamentals} is applied to find the occupancy distribution
at \textit{arbitrary times}, $\piS$, as a function of $\piD$.

In what follows, in order to make a clear distinction between the aforementioned two different steady-state probabilities, we define two different sets of states and a corresponding terminology for their probability distribution as follows:

\begin{itemize}
  \item {\bf queue state at departure instants:} number of packets stored in the queue immediately after a departure. Hereafter, the steady-state probability distribution for these states is referred to as the {\it departure distribution}. This is what was denoted above by the row vector $\piD$.
  \item {\bf queue state at arbitrary times:} number of packets stored in the queue at any arbitrary time. The steady-state probability distribution for these states will simply be referred to as the {\it steady-state distribution} of the queue. This is what was denoted above by the row vector $\piS$.
\end{itemize}

\subsubsection{Departure Distribution, $\piD$}

\begin{figure*}[ttt!!!!!!!!!!!!]

\psfrag{0}[][][0.8]{$0$}
\psfrag{1}[][][0.8]{$1$}
\psfrag{i}[][][0.8]{$i$}
\psfrag{j}[][][0.8]{$j$}

\psfrag{smin}[][][0.8]{$s_{\min}$}
\psfrag{sm}[][][0.8]{$s_{\max}$}
\psfrag{smin-1}[][][0.5]{$s_{\min}-1$}
\psfrag{smax-1}[][][0.5]{$s_{\max}-1$}
\psfrag{K-smin}[][][0.5]{$K-s_{\min}$}
\psfrag{K-smax}[][][0.5]{$K-s_{\max}$}
\psfrag{K-i}[][][0.7]{$K-i$}
\psfrag{K-j}[][][0.7]{$K-j$}

\psfrag{pi0}[][][0.8]{$p_{i,0}$}
\psfrag{pii}[][][0.8]{$p_{i,i}$}
\psfrag{pij}[][][0.8]{$p_{i,j}$}
\psfrag{pismin}[][][0.8]{$p_{i,s_{\min}}$}
\psfrag{pik-smin}[][][0.8]{$p_{i,K-s_{\min}}$}
\psfrag{pik-i}[][][0.8]{$p_{i,K-i}$}

\psfrag{pii-smax}[][][0.8]{$p_{i,i-s_{\max}}$}
\psfrag{pik-smax}[][][0.8]{$p_{i,K-s_{\max}}$}

\centering
\epsfig{file=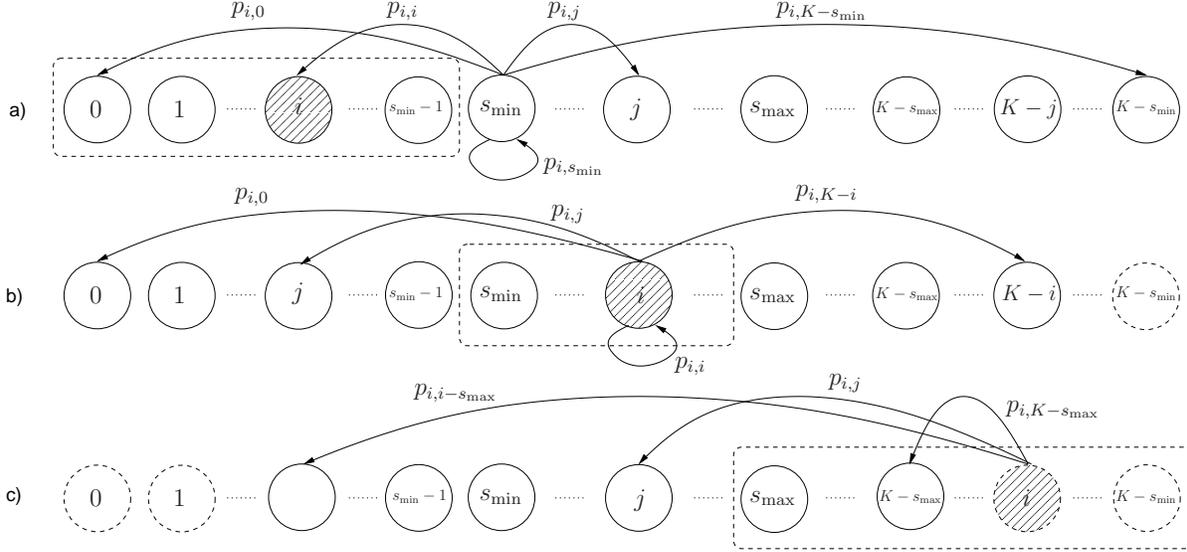,scale=0.45,angle=0}
\caption{Embedded Markov Chain at departure instants for the three possible different regions stated in (\ref{Eq:batch_policy}): dashed states are not reachable after a transmission starting at state $i$.}
\label{Fig:EMC}
\end{figure*}

The departure probability distribution, $\piD$, is computed by solving the
discrete-time embedded Markov chain\footnote{The Markov property of $q_n$ is easily deduced from (\ref{Eq:queue_recursion}).} of the occupancy of the batch-service queue immediately after departure instants, which can be done by solving the linear system $\piD = \piD \Pmat$, together with the normalization condition $\piD \mathbf{1}^{T} = 1$, where $\Pmat$ is the probability transition matrix of the embedded Markov chain, with each element $p_{i,j}$, $i,j\in[0,K-s_{\min}]$, representing the probability to move from state $i$ to state $j$. In this chain, transitions occur at the departure moments, and states represent the queue occupancy immediately after a departure.

The irreducible embedded Markov chain consists of states $i \in[0,K-s_{\min}]$. Note that, even though the queue has a capacity of $K$ packets, a departure can never leave the queue in any state $i>K-s_{\min}$ because at least $s_{\min}$ packets are sent at every departure instant and these packets are held in the queue until their departure. Each transition occurs at a departure instant, which is immediately after receiving the last ACK, for a successful space-batch departure (Figure \ref{Fig:cycles}), or at the moment a packet is discarded, for the packets that reach the retransmission limit. Using a similar reasoning, from any given state $i$, the system can only transition into a subset of states, i.e., $j\in\left[[i-s_{\max}]^+,K-s(i)\right]$.

Figure \ref{Fig:EMC} depicts this Markov chain and the possible transitions from an initial state $i$ for the three possible different cases depending on the value of $i$:
\begin{enumerate}[a)]
 \item When $i< s_{\min}$, the queue does not schedule a new transmission until there are at least $s_{\min}$ packets available. Therefore, $s(i) = s_{\min}$ for any $i< s_{\min}$, and transition to any state $j$ is equivalent to transition from state $s_{\min}$ to state $j$, and therefore, $p_{i,j} = p_{s_{\min},j}$. In this case, the transitions are possible only into states $j \in [0, K-s_{\min}]$.
 \item When $s_{\min}\leq i < s_{\max}$, we have $s(i) = i$ and therefore, transitions from state $i$ can reach any state $j\in\left[0,K-i\right]$.
 \item Finally, when $i\geq s_{\max}$, we have $s(i) = s_{\max}$, and the only possible transitions are into states $j\in[i-s_{\max},K-s_{\max}]$.
\end{enumerate}

The probability of reaching a state $j$ from any state $i$ is given by:
\begin{equation}\label{Eq:transition_pr}
    p_{i,j}=\left \{\begin{array}{lr}
			\pr{V[s(i)] = j - [i-s_{\max}]^+}, & ~ j < K-s(i) \\
			\pr{V[s(i)] \geq j - [i-s_{\max}]^+}, & ~ j = K-s(i)
          \end{array}\right.
\end{equation}
where $j \in \left[[i-s_{\max}]^+,K-s(i)\right]$. For all other values of
$i,j$, we have $p_{i,j} = 0$. Note that $j =K-s(i)$ is the last reachable state from any state $i$. A transition into state $j = K-s(i)$ happens when the queue contains $q = K$ packets (is full) just before the departure, and therefore, some arrivals have possibly been blocked. For all other reachable states, the queue has had room for more packets just before the departure and therefore no arrivals could have been blocked.

The number of arrivals during the transmission of a given space-batch depends on the time required for the transmission of the space-batch, which in turn depends on the space-batch size and the specific channel access protocol considered, as will be discussed in Section \ref{sec:MAC}. Let $X(s(i))$ be the random variable representing the service time for a space-batch of size $s(i)$ packets, and $f_{X(s(i))}(x)$ be the probability density function (pdf) of $X(s(i))$. Note that $X(s(i))$ includes the time required for possible retransmissions due to collisions or errors. For Poisson arrivals of rate $\lambda$, the number of arrivals during $X(s(i))$ has the following distribution:
\begin{equation}\label{Eq:arrival_per_cycle_pdf}
	    \pr{V[s(i)] = v}=\int_{0}^{\infty}{e^{-\lambda t}\frac{(\lambda t)^v}{v!} f_{X(s(i))}(t)}dt
\end{equation}

For any feasible state pair $(i,j)$, i.e., $i\in \left[0,K-s_{\min}\right]$ and $j \in \left[[i-s_{\max}]^+,K-s(i)\right]$, from (\ref{Eq:transition_pr}) and (\ref{Eq:arrival_per_cycle_pdf}), we have:
\begin{equation}\label{Eq:p_ij}
    p_{i,j}=\left \{\begin{array}{lr}
			\displaystyle{\int_{0}^{\infty}{e^{-\lambda
t}\frac{(\lambda
t)^v}{v!} f_{X(s(i))}(t)}dt} ,	& ~ j < K-s(i) \\
			\displaystyle{1-\sum_{z=i-s(i)}^{K-s(i)-1}{p_{i,z}}} ,
& ~ j = K-s(i)
          \end{array}\right.
\end{equation}
where $v = j-[i-s_{\max}]^+$.

Finally, from the departure distribution, the probability to schedule a
space-batch of exactly $s$ packets (at its initial attempt) can be computed as:
\begin{equation}\label{Eq:prob_s_frames}
  \varPsi(s)=\left\{\begin{tabular}{lr}
		$0$ & $s \leq s_{\min}$ \\
		$\sum_{q=0}^{s_{\min}}{\pi^{\text{d}}_q}$ & $s =s_{\min}$ \\
		$\pi^{\text{d}}_s$, & $s \in [s_{\min}+1,s_{\max}-1]$ \\
                $\sum_{q=s_{\max}}^{K-s_{\min}}{\pi^{\text{d}}_q}$ & $s = s_{\max}$ \\
               \end{tabular} \right.
\end{equation}

\subsubsection{Steady-State Distribution, $\piS$}

\begin{figure*}[ttt!!!!!!!!!!!!]
\centering
\epsfig{file=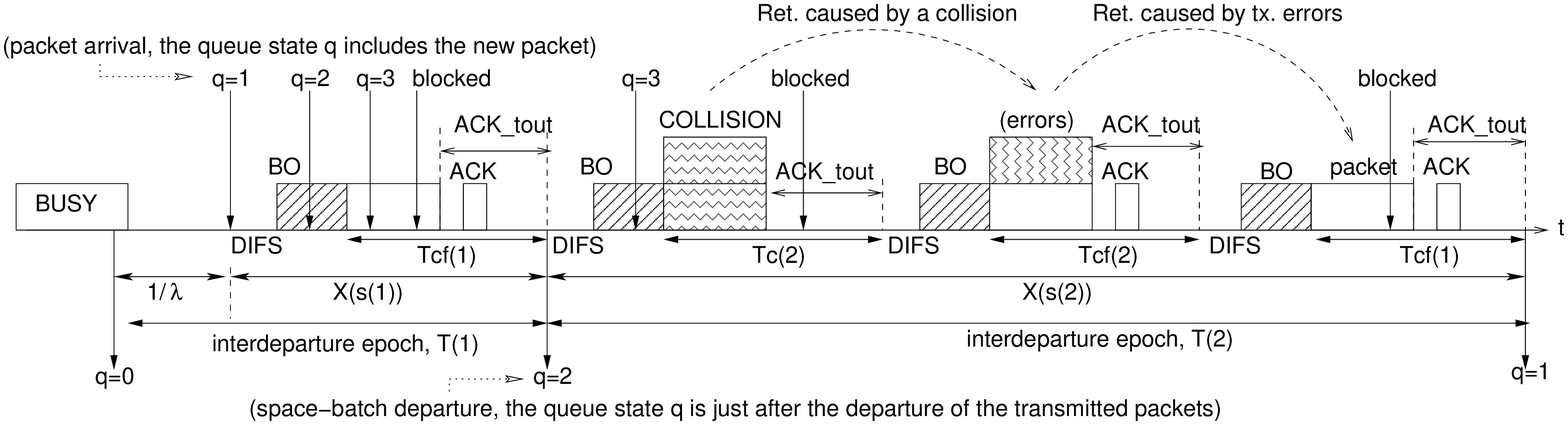,scale=0.475,angle=0}
\caption{Temporal evolution of the queue state for a single node with a queue
length of $K=3$ packets, $M>2$ antennas, $s_{\min}=1$ and $s_{\max}=2$. After the departure of the first space-batch at state $q = 0$, the system remains in the empty state until a new packet arrives, at which instant a new space-batch transmission involving a single packet is initiated. The duration of this interdeparture epoch is $T(0)=\frac{1}{\lambda}+X(s(1))$. The system departs at state $q = 2$ since during $X(1)$, three packets have arrived to the queue.  Notice that the last arrival is blocked as it observes the queue fully occupied ($q = K$). A new space-batch is scheduled immediately at the beginning of the second interdeparture epoch, involving the two packets waiting for transmission. In this case the entire interdeparture epoch is spent in transmission mode and therefore, $T(2)=X(s(2))$. Observe how the second interdeparture epoch includes retransmissions caused by a collision (all packets are retransmitted) and transmission errors (only the non-acknowledged packets are retransmitted).}
\label{Fig:cycles}
\end{figure*}

Using the PASTA property \cite{gross1998fundamentals} of Poisson arrivals, the probability that at an arbitrary time in the steady-state the queue contains $q = k$ packets is equal to the probability that a random arrival observes $k$ packets in the queue. In other words,
\begin{equation} \label{Eq:pis1}
  \pi^{\text{s}}_k=\pr{q_{\text{a}}{(t)}=k}
\end{equation}
where $\pi^{\text{s}}_k$ is the $k$-th element of $\piS$, and $q_{\text{a}}{(t)}$ is the state of the queue observed by an arrival at time $t$. The right hand side of (\ref{Eq:pis1}) can be expanded by conditioning on $q_{\text{d}}(t)$, the state of the queue at the most recent departure before $t$, i.e.,

\begin{equation} \label{Eq:pis2}
  \pi^{\text{s}}_k = \sum_{i=0}^{k}{\pr{q_{\text{a}}{(t)}=k|q_{\text{d}}{(t)}=i}\pr{q_{\text{d}}{(t)}=i}}
\end{equation}

The probability $\pr{q_{\text{d}}{(t)}=i}$, can be viewed as the probability that the arrival at time $t$ happens to occur during the departure state $i$ of the embedded Markov chain discussed in the previous subsection. Therefore, this probability is equal to the expected fraction of time that the node spends in the departure state $i$, i.e.,
\begin{equation} \label{Eq:depart_i}
  \pr{q_{\text{d}}{(t)}=i} = \frac{\pi^{\text{d}}_i E[T(i)]}{\sum_{j=0}^{K}{\pi^{\text{d}}_j E[T(j)]}} = \frac{\pi^{\text{d}}_i E[T(i)]}{E[T]}
\end{equation}
where $\pi^{\text{d}}_j$ is the $j$-th element of $\piD$, the departure distribution, $T(i)$ is the random variable denoting the time spent in departure state $i$, and therefore $E[T]$ is the expected length of an interdeparture epoch.

To calculate $E[T(i)]$, note that of the time spent in departure state $i$, $X(s(i))$ seconds will be spent in transmission mode, and if $i < s_{\min}$, an additional $I(s(i))$ seconds will be spent in idle mode before entering transmission mode. Therefore:
\begin{eqnarray}\label{Eq:epoch_duration}
    E[T(i)] &=& E[I(s(i))] + E[X(s(i))] \nonumber \\
            &=&  \frac{1}{\lambda}\left[ s_{\min}-i \right]^+ + E[X(s(i))]
\end{eqnarray}
where $\frac{1}{\lambda}\left[ s_{\min}-i \right]^+$ is nonzero only when $i<s_{\min}$ and is equal to the expected time needed for the queue occupancy to reach $s_{\min}$. An example of the temporal evolution of the system is depicted in Figure \ref{Fig:cycles}.

The term ${\pr{q_{\text{a}}{(t)}=k|q_{\text{d}}{(t)}=i}}$ in (\ref{Eq:pis2}) is the probability that an arrival during the departure state $i$ observes $k$ packets in the queue, which can be viewed as the expected fraction of arrivals in departure state $i$ that observe $k$ packets in the queue. The expected total number of arrivals in state $i$ is given by $\lambda E[T(i)]$. Of these, only one may observe $k<K$ frames in the queue, provided that there are enough arrivals. Let $q_{n+1}$ be the state at which the next departure will leave the queue. Then the queue occupancy just before this next departure is $q_{n+1}+s(i)$. In order for an arrival to have observed $k$ packets in the queue, we need $q_{n+1} + s(i) \geq k+1$. The expected number of arrivals in state $i$ that observe $k$ packets in the queue is given by:
\begin{equation}
\pr{q_{n+1} \geq k+1-s(i) | q_n = i} = \sum_{j = k+1-s(i)}^{K-s(i)} p_{i,j}
\end{equation}
Based on this, we have
\begin{equation}\label{Eq:prob_i_j_dep}
 {\pr{q_{\text{a}}{(t)}=k|q_{\text{d}}{(t)}=i}} = \frac{\sum_{j = k+1-s(i)}^{K-s(i)} p_{i,j}}{\lambda E[T(i)]}
\end{equation}

From (\ref{Eq:pis2}), (\ref{Eq:depart_i}), and (\ref{Eq:prob_i_j_dep}), the steady state queue occupancy distribution, $\piS$,  for states $0 \leq k \leq K-1$ can be computed as

\begin{equation} \label{Eq:pis_k}
\begin{footnotesize}
  \pi^{\text{s}}_k=\left \{\begin{array}{lr}
			\displaystyle \frac{1}{\lambda E[T]}\sum_{i=0}^{k} \pi_i^{\text{d}} \left(\sum_{j = k+1-s(i)}^{K-s(i)} p_{i,j}\right), & ~ 0\leq k \leq K-1 \\
			\displaystyle 1 - \sum_{i = 0}^{K-1} \pi^{\text{s}}_i, & ~ k = K
          \end{array}\right.
\end{footnotesize}
\end{equation}

\subsection{Characterizing the Service Time}\label{sec:MAC}

As we saw in the previous subsection, in order to find the steady-state queue occupancy distribution, we need to calculate $f_{X(s(i))}(t)$, the probability distribution of the service time when a space-batch of size $s(i)$ packets is transmitted.

However, to keep this subsection as simple as possible, the service time is assumed to follow an exponential distribution, $f_{X(s(i))}=\mu_ie^{-\mu_it}$, where $\mu_i = 1/E[X(s(i))]$. This assumption is done as a tradeoff between accurately capturing the variations in the service time caused by the channel access mechanism, and reducing the numerical complications introduced when solving the model using more accurate distributions (which are different for each case and not known {\it a priori}). Moreover, even though it has been proven that the service time for the DCF in both saturated and non-saturated conditions follows a skewed distribution \cite{zhai2004performance}, the assumption of exponentially distributed service times has been widely used as it provides a reasonable level of accuracy \cite{zhai2004performance}. However, if a more accurate model is required, the specific service time distribution can be computed and included in the model in a similar way as described in \cite{lu2005performance,lu2009performance}. Finally, in the particular case of exponentially distributed $X(s(i))$ with $\mu_i = 1/E[X(s(i))]$, (\ref{Eq:arrival_per_cycle_pdf}) can be simplified to $$\pr{V[s(i)] = v} = \frac{\mu_i}{\mu_i+\lambda}\left(\frac{\lambda}{\mu_i+\lambda}\right)^{v}$$

Once we have assumed that the service time follows an exponential distribution, we only need to obtain $E[X(s)]$, the expected service time for a space-batch of size $s$ packets. Since all the nodes are assumed to have the same behavior, when calculating different MAC layer parameters, we will use this symmetry and focus on a single reference node.

The time required for the successful transmission of a space-batch depends on the number of times it has to be retransmitted due to collisions or transmission errors, on the duration of each transmission, which itself depends on the number of packets sent, and finally, on the random duration of the backoff pauses caused by transmissions from other nodes. Therefore, in order to calculate $E[X(s)]$, we need the probability that the reference node collides with any other node, the packet error probability, the probability that any other node transmits while the reference node is in backoff, and the average duration of the pauses caused by the latter. The probability that a transmitted packet contains errors was previously discussed and is given by (\ref{Eq:PER_SINR}). In this section, we turn our attention to the Medium Access Control to calculate the rest of the aforementioned probabilities.

We adapt the definition of time slots used in \cite{bianchi2000performance}, which defines a slot to be the time in between two decrements of the backoff counter, which is not constant, as the backoff counter is paused every time a transmission is detected on the channel.

\subsubsection{Conditional Collision Probability}

Under saturated conditions, a node will always have packets to transmit, and therefore, after completing every transmission, it will select another random backoff value and start counting down. In this case, the node's backoff counter reaches zero on average every $E[B]+1$ slots, where $E[B]=\frac{CW-1}{2}$ is the expected number of slots a random backoff will last. Therefore, the probability that a node in saturation transmits in a randomly selected slot in steady-state is given by $\tau_{\text{s}}=\frac{1}{E[B]+1}$.

When the network is not saturated, a node can transmit only if it has at least $s_{\min}$ packets ready to be transmitted and its backoff counter has reached $0$ at the end of the previous slot. Therefore, for unsaturated traffic, the probability that during a given backoff slot of the reference node, another given node starts transmitting can be approximated by \cite{tickoo2004queueing}:
\begin{equation}\label{Eq:transmission_probability}
  \tau=\rho \tau_{\text{s}} = \frac{\rho}{E[B]+1}
\end{equation}
where $\rho$ is the steady-state probability that the node has at least $s_{\min}$ packets stored in the queue, $\rho=1-\sum_{i=0}^{s_{\min}-1}{\pi^{\text{s}}_i}$. Note that under saturated conditions, $\rho = 1$ and therefore, (\ref{Eq:transmission_probability}) can also be used for saturated traffic.

Since the nodes sense the channel at the beginning of every backoff slot and pause their backoff if they sense a transmission, a transmission from the reference node will collide only if at least one other node starts transmitting at the same time as the reference node. Therefore, the conditional collision probability, i.e., the probability that a transmission by the reference node ends up in a collision is given by:
\begin{equation}\label{Eq:collision_probability}
  p=1-(1-\tau)^{N-1}
\end{equation}

\subsubsection{Expected Number of Retransmissions}

{Space-batch retransmissions are caused by collisions or transmission errors. In case of a collision, all packets included in the space-batch are lost and should be retransmitted. In case of a collision-free transmission containing $m$ packets, from $0$ to $m$ packets can contain errors, each one with probability $\PER(m)$, independently of the rest. In this case, only the unacknowledged packets will be retransmitted. Let $\Upsilon_{\text{cf}}(s)$ be the average number of collision-free transmission attempts required to successfully transmit all the $s$ packets included in the initial space-batch, which includes all the erroneous transmission attempts and a single successful one. Let $\Upsilon_{\text{c}}$ be the average number of transmission attempts required for every collision-free transmission, i.e., the number of collisions before every collision-free transmission plus the collision-free transmission itself. For example, CE$|$E$|$CCE$|$CS is a possible transmission attempt sequence outcome for transmitting a given space-batch, with C, E, and S indicating  collided, erroneous, and successful transmission, respectively. In this example, the number of collision-free (but possibly erroneous) transmission attempts is $4$. The last collision-free transmission attempt is the only one that is successful. Then, the total expected number of transmission attempts required per space-batch is $\Upsilon(s)=\Upsilon_{\text{cf}}(s) \Upsilon_{\text{c}}$.

Assuming that the collision probability, $p$, remains constant regardless of the number of collisions that the target space-batch has already suffered, and that the maximum number of allowed retransmission attempts per space-batch is infinite, the number of attempts per collision-free transmission follows a geometric distribution, and its average value, $\Upsilon_{\text{c}}$ is given by:
\begin{equation}\label{Eq:number_of_retransmissions_col}
  \Upsilon_{\text{c}}=\sum_{k=1}^{\infty}{k(1-p)p^{k-1}} =\frac{1}{1-p}
\end{equation}

\begin{figure}[t!!!!!!!!]

\psfrag{s}[][][0.8]{s}
\psfrag{s-1}[][][0.8]{$s-1$}
\psfrag{m}[][][0.8]{$m$}
\psfrag{1}[][][0.8]{$1$}
\psfrag{0}[][][0.8]{$0$}

\psfrag{pss}{$\hat{p}_{s,s}$}
\psfrag{ps1ps1}{$\hat{p}_{s-1,s-1}$}
\psfrag{pmm}{$\hat{p}_{m,m}$}
\psfrag{pm1}{$\hat{p}_{m,1}$}
\psfrag{p11}{$\hat{p}_{1,1}$}
\psfrag{pss1}{$\hat{p}_{s,s-1}$}
\psfrag{psm}{$\hat{p}_{s,m}$}
\psfrag{ps1}{$\hat{p}_{s,1}$}
\psfrag{ps0}{$\hat{p}_{s,0}$}
\psfrag{pm0}{$\hat{p}_{m,0}$}
\centering
\epsfig{file=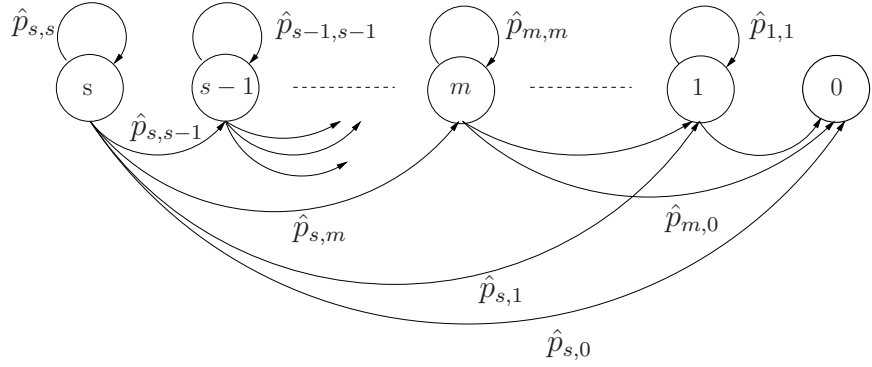,scale=0.45,angle=0}
\caption{Absorbing Markov chain to compute the number of required transmissions due to errors to successfully transmit all the packets included in a space-batch} \label{Fig:AbsorvMC}
\end{figure}

To compute $\Upsilon_{\text{cf}}(s)$, we need to keep track of the unacknowledged packets remaining for retransmission, for which we use an absorbing, time-homogeneous Markov chain (Figure \ref{Fig:AbsorvMC}), whose state space is given by $\mathbf{S} =\{m\}_{m = 0}^s$. Each state $m$ represents the remaining number of unacknowledged packets, and $s$, the original size of the space-batch. The Markov chain reaches the absorbing state $m=0$ when all the packets have been successfully transmitted. Transition from a state $i \leq s$ to a state $j \leq i$ requires the successful transmission of $i-j$ of the $i$ packets transmitted in state $i$, which happens with probability: }
\begin{equation}\label{Eq:E_probs}
	\hat{p}_{i,j}=\binom{i}{i-j}(1-\PER(i))^{i-j}\PER(i)^j
\end{equation}

Let $\Phat$ be the transition probability matrix of the chain, with $\hat{p}_{i,j}$ at its $(i,j)$ element. Given that the state $(0,0)$ is absorbing, the first row of $\Phat$ contains all zeros except for its first element which is one. In this case, $\Phat$ can be represented in the following canonical form:

\begin{equation} \label{Eq:AMC_canonical}
\Phat \;= \bordermatrix{
                       &\hbox{ABS}  &\omit\hfil &\hbox{TR}\cr
    \hbox{ABS}\bigstrut &\mathbf{1} &\srule     &\mathbf{0}_{(1 \times s)} \cr
\middlehrule{1}{1}
    \hbox{TR}\bigstrut&\mathbf{R}_{(s \times 1)} &\srule &\mathbf{Q}_{(s\times s)}}
\end{equation}
where the rows and columns corresponding to the transient and absorbing states are indicated by TR and ABS, respectively. For an absorbing chain with the canonical representation above, the following properties hold \cite{grinstead1997introduction}:

\begin{enumerate}
\item The matrix \mbox{$\mathbf{I} - \mathbf{Q}$} has an inverse given by $\mathbf{N}  =\mathbf{I} + \mathbf{Q} + \mathbf{Q}^{2} + \cdots$, with each of its elements, $n_{ij}$, being the expected number of times the chain visits state $j$ before absorption, given that it started in state $i$. 

\item Let $\Upsilon_{\text{cf}}(i)$ be the expected number of transitions before absorption, starting from state $i$, and $\mathbf{\Upsilon_{\text{cf}}}$ a column vector whose $i$-th entry is $\Upsilon_{\text{cf}}(i)$. Then $\mathbf{\Upsilon_{\text{cf}}} = \mathbf{N}\cdot\mathbf{c}$, where $\mathbf{c}$ is an $s \times 1$ column vector containing all ones.
\end{enumerate}

The number of required transmissions to successfully transmit all packets in a space-batch initially containing $s$ packets, $\Upsilon_{\text{cf}}(s)$, is equal to the expected number of steps to absorption in the aforementioned absorbing Markov chain, starting at state $s$. By solving the chain only once for $s=M$, we can obtain all $\Upsilon_{\text{cf}}(i)$ values for all $i$.

\subsubsection{Expected Service Time}

The expected service time for a space-batch of size $s$, $E[X(s)]$, is the time since it is scheduled until either the successful arrival of all of its packets is confirmed, or it is discarded due to reaching the maximum number of retransmissions, and is given by:
\begin{equation}\label{Eq:service_time}
 E[X(s)]= \Upsilon(s) E[B]\gamma + \Upsilon_{\text{cf}}(s)\left(\left(\overline{T}_{\text{cf}}(s)+\DIFS+\sigma\right)+(\Upsilon_{\text{c}}-1)\left(\overline{T}_{\text{c}}(s)+\DIFS+\sigma\right)\right)
\end{equation}
where $\gamma$ is the average duration of a backoff slot and includes possible backoff countdown interruptions, and $\overline{T}_{\text{cf}}(s)$ and $\overline{T}_{\text{c}}(s)$ are the expected duration of  collision-free and collided transmission attempts, respectively, during the transmission of a space-batch of initial size $s$, which will be computed shortly. The extra empty slot duration ($\sigma$) and the $\DIFS$ added to $\overline{T}_{\text{cf}}(s)$ and $\overline{T}_{\text{s}}(s)$ respectively account for the fact that after a transmission, the reference node waits for an extra empty slot before restarting its backoff counter and that the channel is only detected idle after sensing it free for $\DIFS$ seconds. The first term in the right-hand-side of (\ref{Eq:service_time}), represents the total amount of time spent in backoff over all transmission attempts required to send a space-batch of size $s$, and the second term the time spent in transmission and receiving acknowledgements. The expected service time averaged over the space-batch size, $s$, is simply

\begin{equation}
	E[X]=\sum_{s=s_{\min}}^{s_{\max}}{\varPsi(s)E[X(s)]}
\end{equation}

As mentioned before, the duration of a slot can vary depending on whether or not at the beginning of that slot a transmission is detected on the channel. Then the average slot duration, $\gamma$, can be calculated as:
\begin{equation}\label{Eq:gamma}
  \gamma = p_{\text{e}} \sigma + p_{\text{cf}} (\overline{T}_{\text{cf}}+\DIFS+\sigma) + p_{\text{c}} (\overline{T}_{\text{c}}+\DIFS+\sigma)
\end{equation}
where $p_{\text{e}}$, $p_{\text{cf}}$, and $p_{\text{c}}$ are the probabilities that during a given backoff slot of the reference node the channel is empty, contains a collision-free transmission, or contains a collision, respectively. $\overline{T}_{\text{cf}}$ and $\overline{T}_{\text{c}}$ are the expected duration of a collision-free and a collided transmission observed by the reference node in backoff, as given by equations (\ref{Eq:ETcf}) and (\ref{Eq:ETs_av}).

A node performing backoff will find the channel idle if there is no other node transmitting at that slot. Therefore, the probability $p_{\text{e}}$, that a given node finds the channel empty during a backoff slot is given by:
\begin{equation}\label{Eq:ch_empty}
  p_{\text{e}} = \left(1-\tau \right)^{N-1}
\end{equation}

The probability of finding a successful transmission during a backoff slot, $p_{\text{cf}}$, is equal to the probability that only one of the other $N-1$ active nodes transmit during that slot, i.e.,
\begin{equation}\label{Eq:ch_success}
  p_{\text{cf}} = (N-1) \tau\left(1-\tau \right)^{N-2}
\end{equation}

Finally, the probability that a selected slot contains a collision, $p_{\text{c}}$ is the complementary of the two previous cases
\begin{equation}\label{Eq:ch_collision}
  p_{\text{c}} = 1-p_{\text{cf}}-p_{\text{e}}
\end{equation}

{To compute $\overline{T}_{\text{cf}}$ and $\overline{T}_{\text{c}}$ it should be noted that both are expected durations averaged over all possible values of $m$, the number of packets included in each transmission. The number of packets included in a transmission attempt depends on the initial number of packets, $s$, scheduled for that space-batch and the number of packets not successfully transmitted in the previous attempts. Let $p_{m|s}$ denote the probability that a given transmission attempt during the transmission of a space-batch of initial size $s$, happens to contain $m$ packets. Then, the average duration for each transmission attempt for a space-batch of initially $s$ packets, $\overline{T}_{\text{cf}}(s)$, is given by:

\begin{equation}\label{Eq:ETcf_s}
  \overline{T}_{\text{cf}}(s)={\sum_{m=1}^{s}{p_{m|s}T_{\text{cf}}(m)}}
\end{equation}
where $T_{\text{cf}}(m)$ is given by (\ref{Eq:tx_succ_duration}). Averaging for  different values of $s$ results:

\begin{equation}\label{Eq:ETcf}
  \overline{T}_{\text{cf}}=\sum_{s=s_{\min}}^{s_{\max}}{\varPsi(s)\overline{T}_{\text{cf}}(s)}
\end{equation}

Using a similar approach, for $\overline{T}_{\text{c}}$, we first calculate the duration of a collision given that one of the two nodes involved in the collision has transmitted a space-batch of initially $s$ frames:

\begin{equation}\label{Eq:ETc_s}
  \overline{T}_{\text{c}}(s)\approx \sum_{s_1=s_{\min}}^{s_{\max}}{\varPsi(s_1)\sum_{i=1}^{s_1}{p_{i|s_1}\sum_{j=1}^{s}{p_{j|s}\max(T_{\text{cf}}(i),T_{\text{cf}}(j))}}}
\end{equation}
which is simplified, relying on the fact that the probability of having more than two nodes involved in a single collision is negligibly small. Averaging over all possible $s$ values, we get:

\begin{equation}\label{Eq:ETs_av}
  \overline{T}_{\text{c}} = \sum_{s=s_{\min}}^{s_{\max}}{\varPsi(s)\overline{T}_{\text{c}}(s)}
\end{equation}

The conditional probability of transmitting a space-batch containing $m$ packets when the initial space-batch included $s$ packets, $p_{m|s}$, is calculated using a Markov chain with state space  $\mathbf{\tilde{S}} =\{m\}_{m=1}^s$, where $s$ is the size of the initial space-batch. This Markov chain is exactly the same as the absorbing Markov chain used to calculate $\Upsilon_{\text{cf}}(s)$, except that, in order to be able to compute the stationary distribution, the absorbing state has been removed. In this case, when all packets have been successfully transmitted, the chain goes back to the initial state $s$, accounting for the next scheduled space-batch of initial size $s$.

Let $\tilde{p}_{i,j}(s)$ be the probability to move from a state $i$ to a state $j$ when the initial space-batch includes $s$ packets, given by:

\begin{equation}\label{Eq:prob_trans_s_frames}
  \tilde{p}_{i,j}(s)=\left\{\begin{tabular}{lr}
		$\hat{p}_{i,j}$ & $1 \leq i \leq s$, $1 \leq j \leq i$ \\
    	$\hat{p}_{i,0}$ & $1 \leq i < s$, $j=s$ \\
    	$\hat{p}_{s,0}+ \hat{p}_{s,s}$ & $i=s$, $j=s$ \\    	
               \end{tabular} \right.
\end{equation}
where $\hat{p}_{i,j}$ are given by (\ref{Eq:E_probs}).

Let $\Ptilde_s$ be the transition matrix of this Markov chain (Figure \ref{Fig:ErrorMC}) and $\pitildeS$, the vector containing its stationary probabilities. Then, $\pitildeS$ is obtained by solving the equation system $\pitildeS = \pitildeS \Ptilde_s$, together with the normalization condition $\pitildeS \mathbf{1}^{T} = 1$. Then, $p_{m|s}$ is the $m$-th element of $\pitildeS$, i.e., $p_{m|s} = \pitildeS (m)$.

\begin{figure}[t!!!!!!!!]
\centering

\psfrag{s}[][][0.8]{s}
\psfrag{s-1}[][][0.8]{$s-1$}
\psfrag{m}[][][0.8]{$m$}
\psfrag{1}[][][0.8]{$1$}

\psfrag{pss}{$\tilde{p}_{s,s}(s)$}
\psfrag{p1s}{$\tilde{p}_{1,s}(s)$}
\psfrag{ps1s1}{$\tilde{p}_{s-1,s-1}(s)$}
\psfrag{pmm}{$\tilde{p}_{m,m}(s)$}
\psfrag{p11}{$\tilde{p}_{1,1}(s)$}
\psfrag{psm}{$\tilde{p}_{s,m}(s)$}
\psfrag{ps1}{$\tilde{p}_{s,1}(s)$}
\psfrag{pm1}{$\tilde{p}_{m,1}(s)$}

\epsfig{file=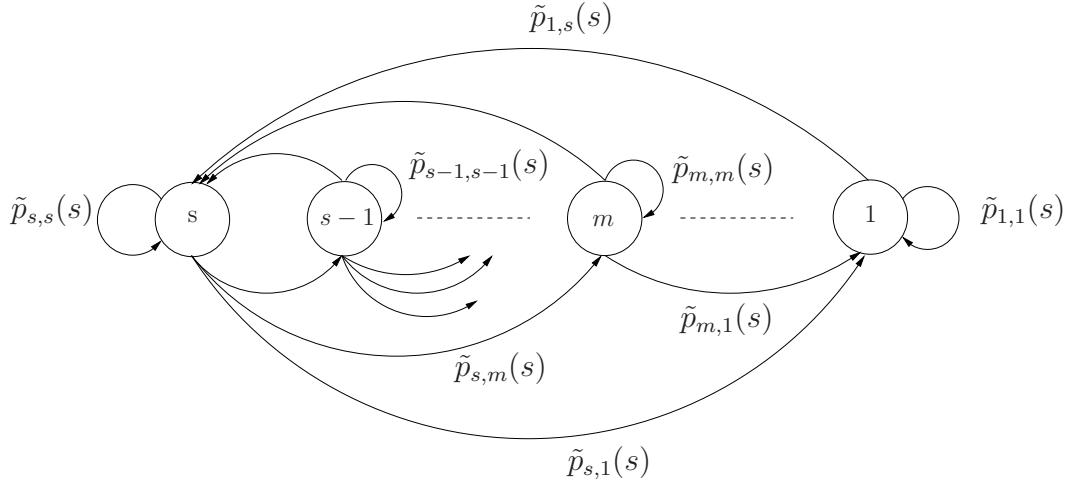,scale=0.5,angle=0}
\caption{Discrete Time Markov Chain to compute the distribution of the duration of
the channel transmissions in presence of transmission errors} \label{Fig:ErrorMC}
\end{figure}

}

%
%

\section{Performance Evaluation} \label{sec:Results}

In this section, the performance of a network formed by $N$ low-complexity wireless objects operating under the reference protocol is evaluated through simulation and is compared to the results obtained by the analytical formulation presented in the previous sections.

\begin{small}
\begin{table}
\centering
\caption{Specific parameters considered for the reference protocol's space-batch and ACK frames}
 \begin{tabular}{c|c||c|c}
  Field & Length & Field & Length\\
  \hline
  PHY header & $192$ bits & MAC header & $160$ bits \\
  Training Sequence (each one) & $64$ bits & Packet length & variable bits\\
 \end{tabular}
  \label{Tbl:frame_fields}
\end{table}
\end{small}

\subsection{Scenario, Simulation, and Model Parameters and Considerations}

In order to evaluate the performance of the reference MAC protocol, a simulator that implements the scenario described in Section \ref{sec:Scenario} has been built in C++, from scratch, based on the COST (Component Oriented Simulation Toolkit) libraries \cite{chen2001component}. Since the simulator reproduces the full operation of the reference MAC protocol, we can validate the accuracy of the approximations made to build the analytical model by comparing the results obtained using each method.

The data transmission rate and the physical transmission rate are set to $r=500$ and $r_{\text{phy}}=250$ Kbps respectively, and a constant packet length of $L=4000$ bits are used. Additionally, for the MAC parameters, a single-stage backoff with $CW=32$, $\DIFS=50~\mu$s, $\SIFS=10~\mu$s, and $\sigma=20~\mu$s is considered. Each node has a queue of size $K=50$ packets. In Table \ref{Tbl:frame_fields}, the values for different headers of the reference protocol's space-batch and ACK frames are shown. With respect to the {\text{PER}}, we have considered a reference SNR of $\xi_{\REF} = 15$ dBs.

Two performance metrics are considered for the evaluation of the MAC protocol:

\begin{itemize}
\item \textit{Aggregate throughput} (packets/second) is defined as the number of packets successfully transmitted in the network per unit of time and is given by $S=N (1-p_{\text{b}})\lambda$, where $p_{\text{b}}=\pi^{\text{s}}_{K}$ is the blocking probability, i.e., the probability that a packet is lost due to finding a full queue upon its arrival.
\item \textit{Expected queueing delay} (seconds) is defined as the amount of time a packet spends waiting in the queue, from the moment it is admitted to the queue until it is purged from the queue after being successfully transmitted {or discarded}. Using Little's law \cite{gross1998fundamentals}, the expected queueing delay is given by $E[D]=\frac{E[Q]}{\lambda(1-p_{\text{b}})}$, where $E[Q]=\sum_{q=0}^{K}{q \pi^{\text{s}}_q}$, is the expected queue occupancy.
\end{itemize}

\subsection{Results}

\begin{figure*}[ttttttttttttt!!!!!]
\begin{center}
\subfigure[]{\epsfig{file=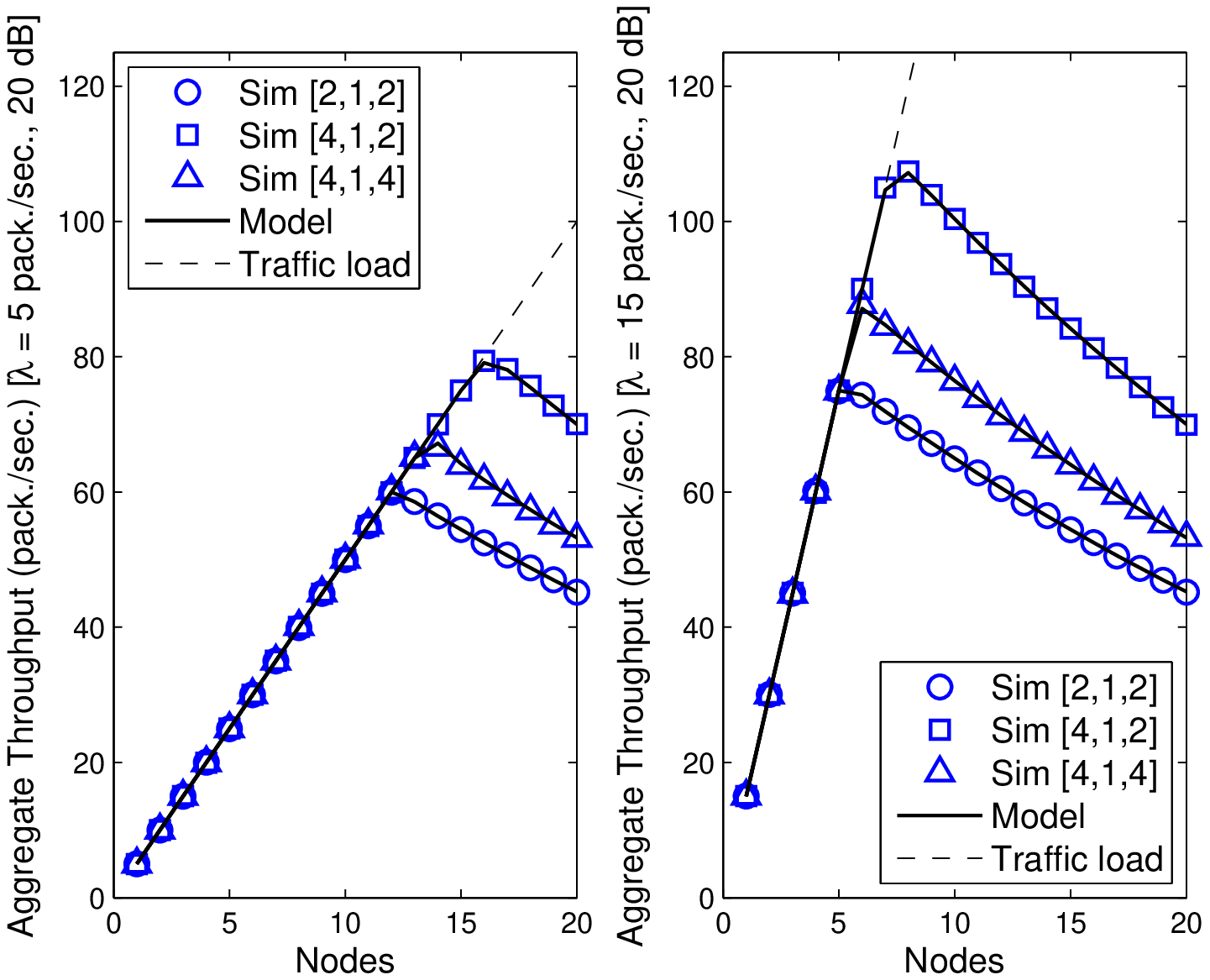,scale=0.45,angle=0}\label{Fig:MAperformance_a}}
\subfigure[]{\epsfig{file=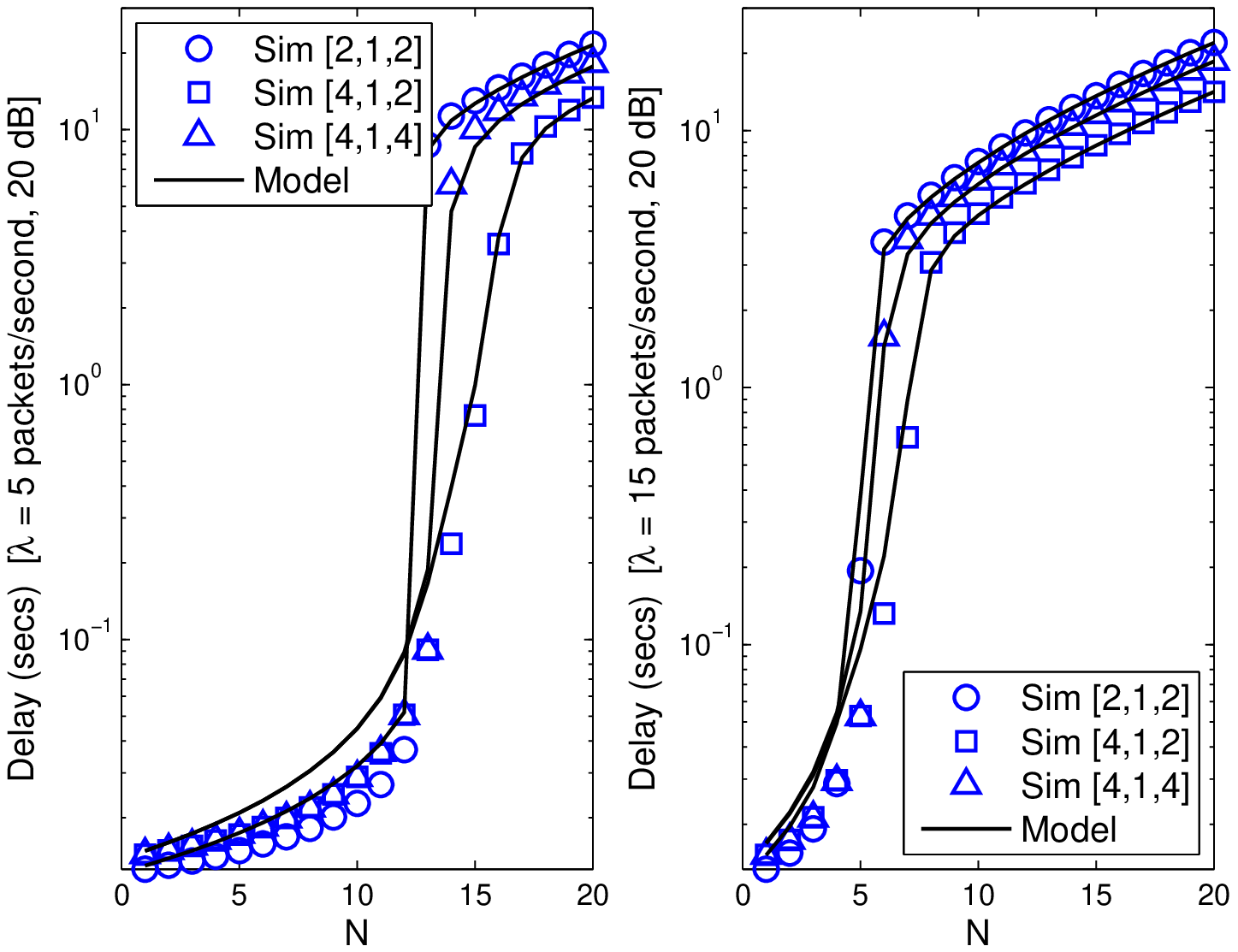,scale=0.45,angle=0}\label{Fig:MAperformance_b}}\\
\subfigure[]{\epsfig{file=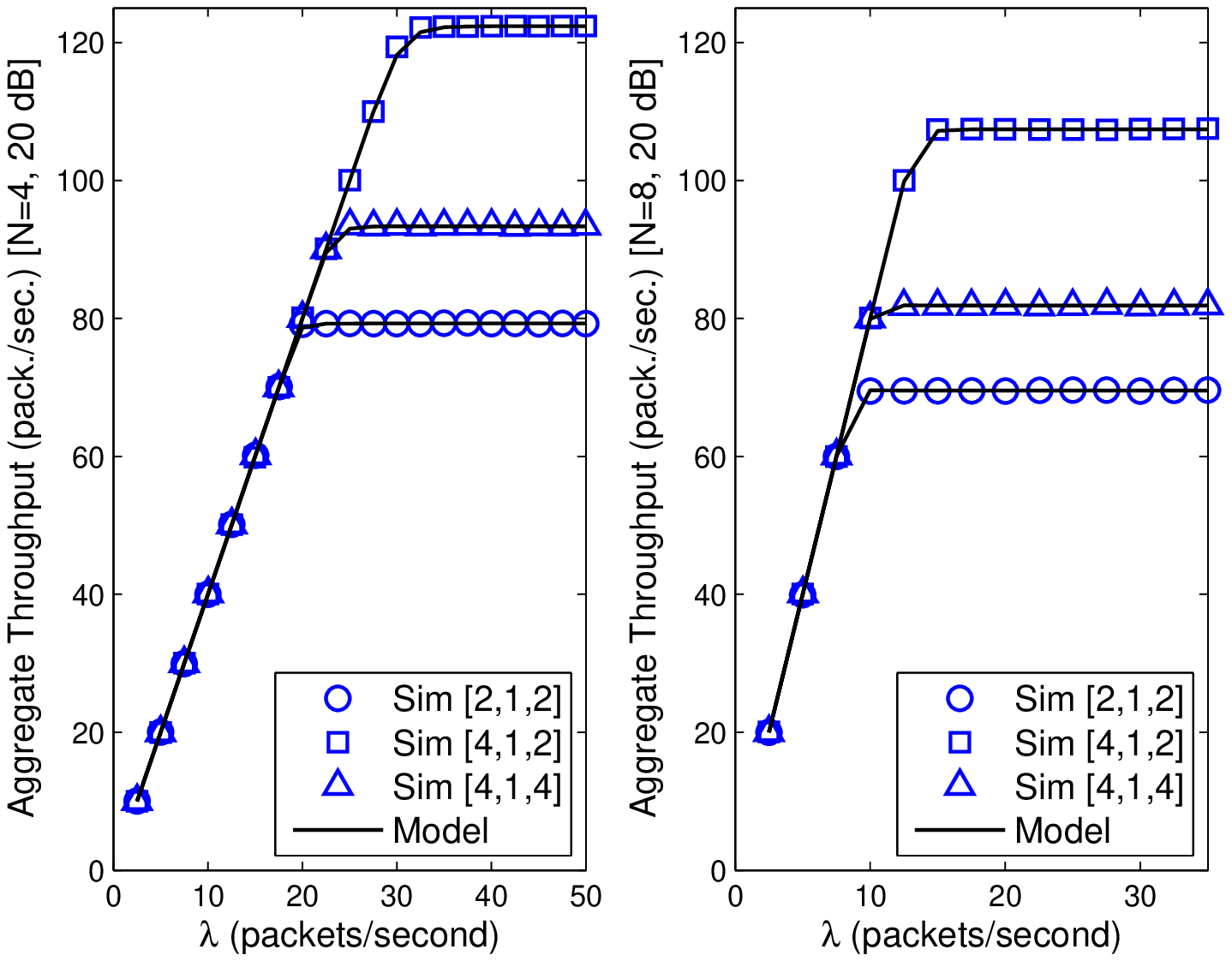,scale=0.45,angle=0}\label{Fig:MAperformance_c}}
\subfigure[]{\epsfig{file=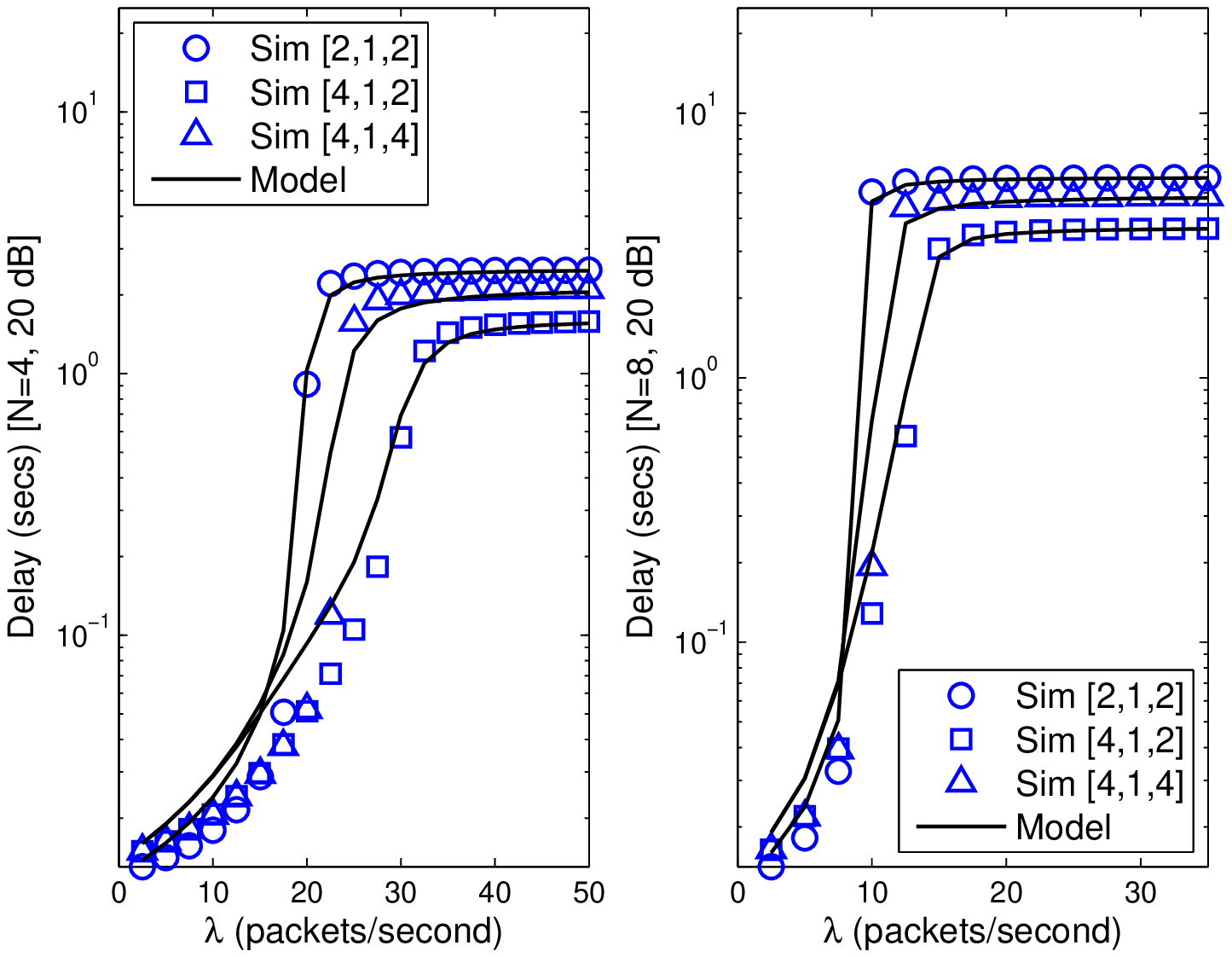,scale=0.45,angle=0}\label{Fig:MAperformance_d}}
\caption{The impact of $M$ and $s_{max}$ on the performance of reference MAC protocol in non-saturation conditions. A SNR value ($\xi_0$) of $20$ dB is considered. In the legend, each curve is marked by the value of $[M,s_{\min}, s_{\max}]$. }\label{Fig:MAperformance}
\end{center}
\end{figure*}

Figures \ref{Fig:MAperformance_a} and \ref{Fig:MAperformance_b} show the aggregate throughput and expected delay, respectively, each for two different arrival rates ($5$ and $15$ packets/second) when $\xi_0=20$ dB, $s_{\min}=1$, $s_{\max}=2$, and $M=2$ or $4$ antennas\footnote{Typical values for the number of antennas in IEEE 802.11-like technologies range between $2$ and $8$, with 8 the maximum number of antennas that an IEEE 802.11ac Access Point will support \cite{IEEE80211ac}.}, as indicated by $[M,s_{\min}, s_{\max}]$ in the legend. For both metrics, the best performance is observed when the nodes are equipped with $M=4$ antennas and $s_{max}$ is set to $2$, showing that in some cases the lower \PER~ that the Zero Forcing receiver is able to provide by using the channel's extra degrees of freedom for diversity, compensates the transmission of less packets at each channel attempt. A similar result is observed in both the aggregate throughput (Figure \ref{Fig:MAperformance_c}) and the expected delay (Figure \ref{Fig:MAperformance_d}) when the number of nodes is fixed to $N=4$ and $8$ nodes, and the packet arrival rate at each node is increased.

\begin{figure*}[tttttttttt!!!!!]
\begin{center}
\subfigure[]{\epsfig{file=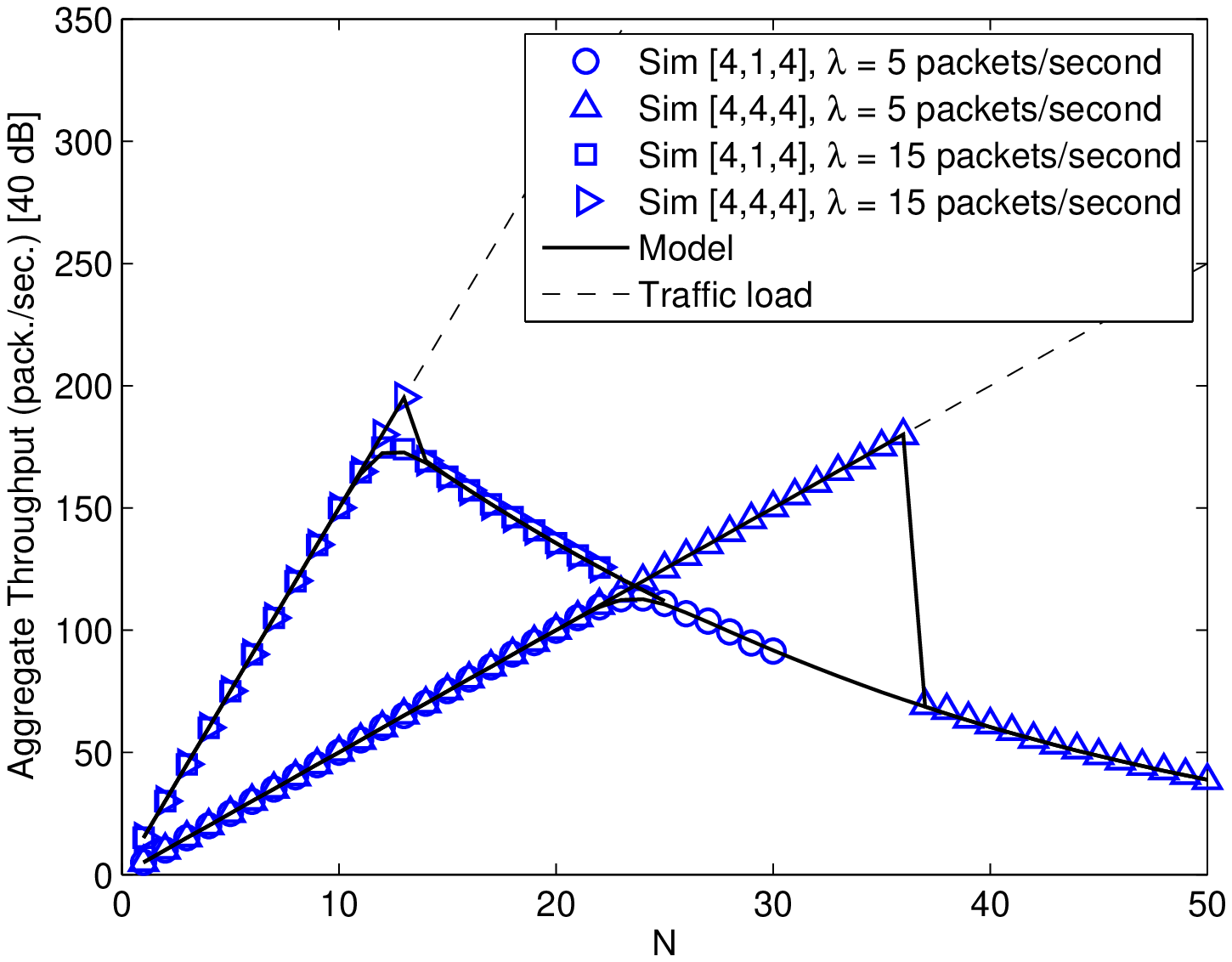,scale=0.45,angle=0}\label{Fig:MAconfiguration_a}}
\subfigure[]{\epsfig{file=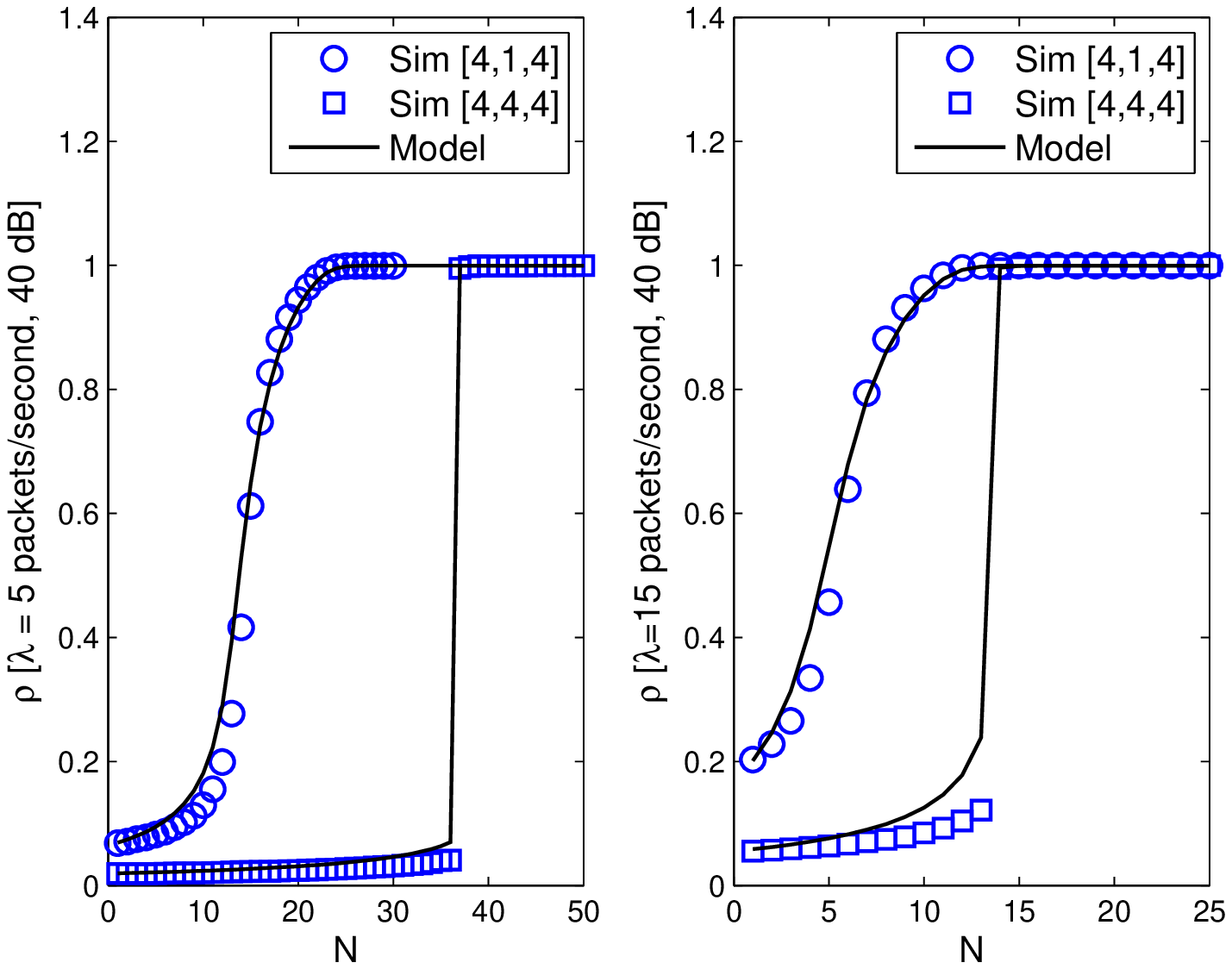,scale=0.45,angle=0}\label{Fig:MAconfiguration_c}}\\
\subfigure[]{\epsfig{file=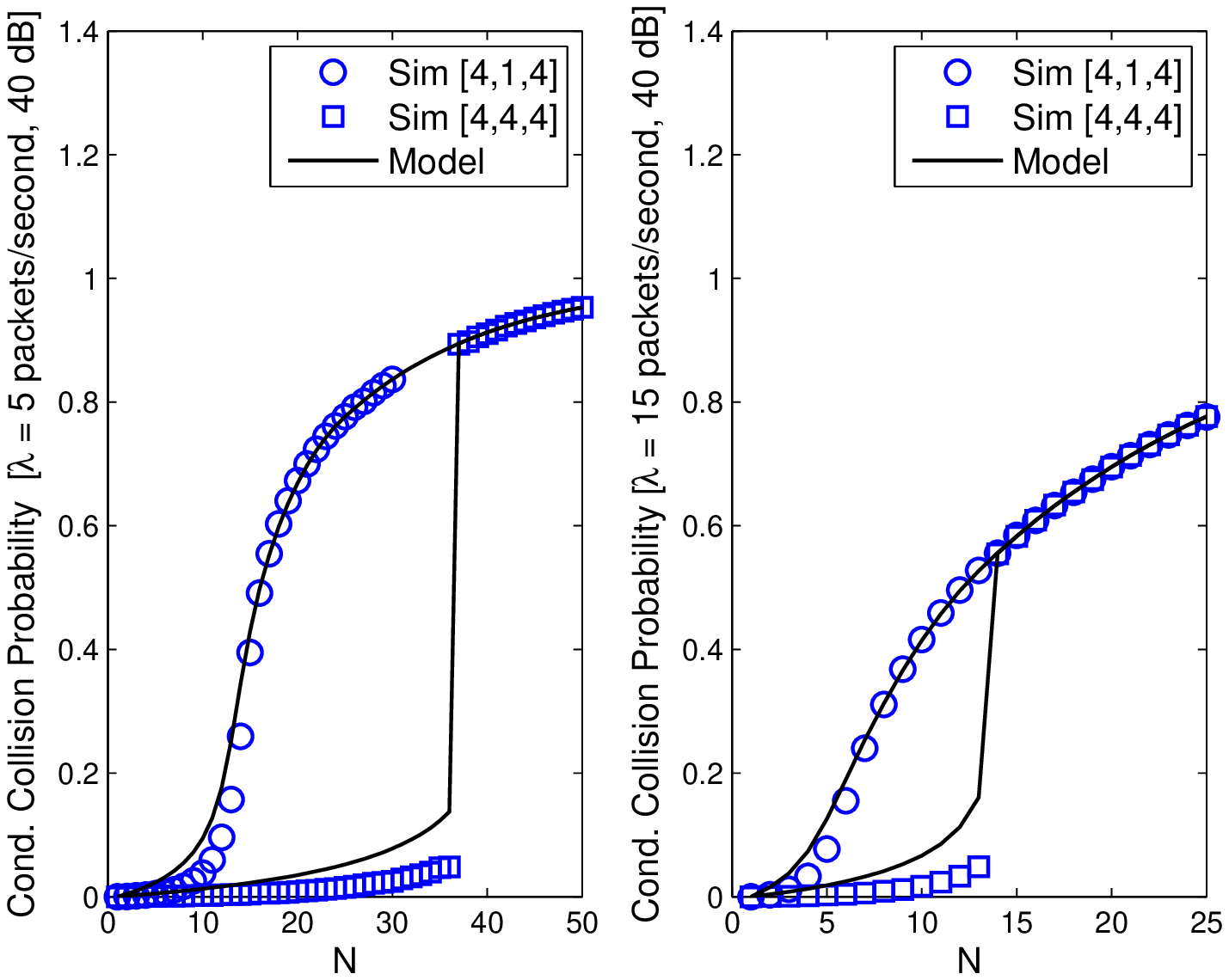,scale=0.45,angle=0}\label{Fig:MAconfiguration_d}}
\subfigure[]{\epsfig{file=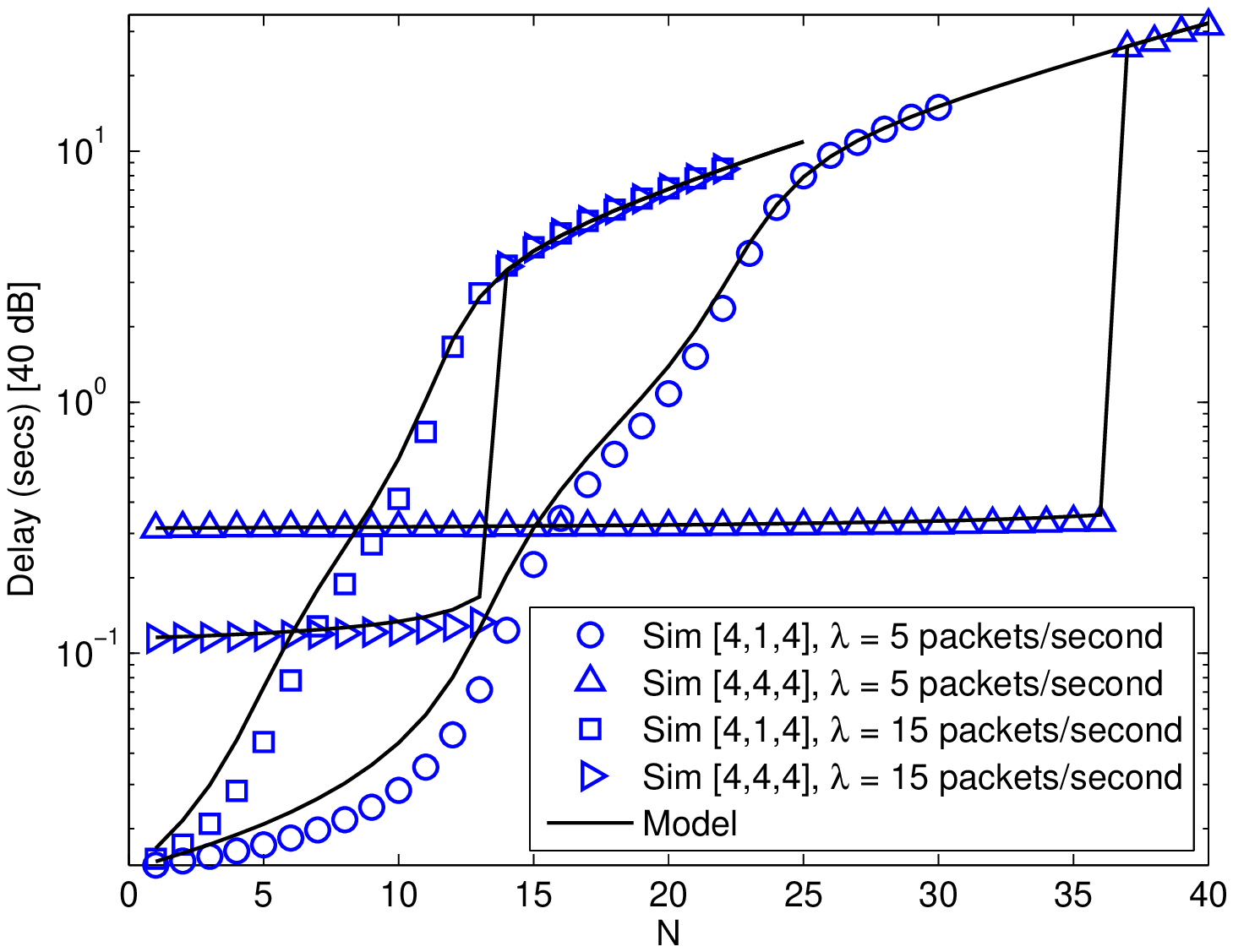,scale=0.45,angle=0}\label{Fig:MAconfiguration_b}}
\caption{{The impact of $s_{\min}$ on the performance of the system in non-saturation conditions. In the legend, each curve is marked by the value of $[M,s_{\min}, s_{\max}]$. A SNR value ($\xi_0$) of $40$ dBs is considered. The packet arrival rate ($\lambda$) of each node is $5$ or $15$ packets/second.}}\label{Fig:MAconfiguration}
\end{center}
\end{figure*}

The impact of different $s_{\min}$ values is evaluated when multiple nodes contend for the channel. The results are plotted in Figure \ref{Fig:MAconfiguration}. The considered parameters are $M=4$ antennas, $\xi_0=40$ dBs, $s_{\max}=4$, and $s_{\min}=1$ or $4$ packets. In this case, in order to highlight the effect of increasing $s_{\min}$ on network performance, the average SNR has been set to $40$ dBs, as opposed to $20$dBs in Figure \ref{Fig:MAperformance} where transmissions containing $4$ packets resulted in high error probability. By increasing the SNR to $40$ dBs in Figure \ref{Fig:MAconfiguration}, the transmission of large space-batches becomes beneficial and setting $s_{\min} = 4$ can be advantageous. As can be observed, setting $s_{\min}=4$ improves the performance of the MAC protocol for larger networks when the traffic load of each node is low (for instance, between $22$ and $36$ nodes for $\lambda=5$ packets/second), compared to the case with $s_{\min}=1$ (Figure \ref{Fig:MAconfiguration_a}). Additionally, an interesting issue here, already known from the analysis of the DCF with single-antenna nodes \cite{malone2007modeling}, is that in some specific circumstances (i.e., when the collision probability just before saturation remains very low), the non-saturated throughput just before the saturation point is higher than the saturation throughput, as it is also observed here. Setting $s_{\min}=4$ results in fewer channel attempts as the queue remains inactive, i.e., with less than $s_{\min}$ packets, for longer periods of time (Figure \ref{Fig:MAconfiguration_c}), which results in a lower collision probability until the saturation point is reached (Figure \ref{Fig:MAconfiguration_d}). However, values of $s_{\min}$ larger than $1$ have a negative impact on the delay (Figure \ref{Fig:MAconfiguration_b}), due to the extra time required to have $s_{\min}$ packets in the queue at low traffic loads (small $N$). Note that this time is proportional to $\frac{1}{s_{\min}}\sum_{i=1}^{s_{\min}}{\frac{1}{\lambda}(s_{\min}-i)}$ seconds.

In all of the plots, the model and the simulation curves show a very good match, with the model always slightly above the simulation curve. This may be justified by the higher randomness of the exponential distribution, as well as by the assumption that the transmission probability at each slot is constant and does not depend on the instantaneous number of nodes that are contending for the channel access, which has been used to obtain the expected service time. However, from the results we can conclude that the model is accurate enough to be useful as a tool for understanding the effect of different scenarios and parameter configurations, as well as for finding robust parameter values that yield to a close-to-optimum performance if the model is embedded in optimization algorithms.

%
%

\section{Conclusions} \label{sec:Conclusions}

In this paper, an analytical model for a CSMA/CA MAC protocol with MPT capabilities has been presented. Simulation results show that the presented model is able to capture the fundamental interactions between the CSMA/CA channel access mechanism and the MPT capabilities of the reference protocol, as well as providing a good approximation of the system performance.

As observed, the required extra overheads to support multiple packet transmission are compensated by the larger amount of data that can be sent at each transmission, as well as by the reduction in the number of transmissions, which in turn results in lower collision probabilities for a given number of contending nodes and a given traffic load, thus providing a higher throughput and lower delay.

Moreover, given a node equipped with multiple antennas, two new parameters called $s_{\min}$ and $s_{\max}$, which control the number of packets that are scheduled at each transmission, were introduced and their impact on the system performance was evaluated, drawing several conclusions. {Mainly, it was observed that, although the number of packets that can be included in a space-batch increases when $s_{\max}$ increases, the \PER~also increases. Therefore, the value of $s_{\max}$ has to be set to the highest possible value that still guarantees a reasonably low \PER~(i.e., based on the observed SNR and available number of antennas in each node).} It was also observed that adapting $s_{\min}$ to the traffic load and the number of active nodes, the system performance can be optimized in terms of delay and throughput. Additionally, although not analyzed in this paper, the impact of both parameters also depends on the backoff algorithm and its parameters (i.e., $CW$ and number of stages), thus opening a new opportunity to search optimal configurations for all those parameters simultaneously. 

%
%

\section{Acknowledgments} \label{sec:ACKs}

This work has been partially supported by COST (European Cooperation in Science and Technology) Action IC0906, the Spanish Government under projects TEC2012-32354, TEC2012-34642 and TEC2009-13000 (Plan Nacional I+D), CSD2008-00010 (Consolider-Ingenio) and by the Catalan Government (SGR2009\#00617 and SGR2009\#70). We would also like to express our gratitude to the reviewers for their insightful comments.



\bibliographystyle{elsarticle-num}
\bibliography{MACMA}

\begin{thebibliography}{10}
\expandafter\ifx\csname url\endcsname\relax
  \def\url#1{\texttt{#1}}\fi
\expandafter\ifx\csname urlprefix\endcsname\relax\def\urlprefix{URL }\fi
\expandafter\ifx\csname href\endcsname\relax
  \def\href#1#2{#2} \def\path#1{#1}\fi

\bibitem{mietzner2009multiple}
J.~Mietzner, R.~Schober, L.~Lampe, W.~Gerstacker, P.~Hoeher, {Multiple-antenna
  Techniques for Wireless Communications-A Comprehensive Literature Survey},
  Communications Surveys \& Tutorials, IEEE 11~(2) (2009) 87--105.

\bibitem{zheng2003diversity}
L.~Zheng, D.~Tse, {Diversity and Multiplexing: A Fundamental Tradeoff in
  Multiple-Antenna Channels}, Information Theory, IEEE Transactions on 49~(5)
  (2003) 1073--1096.

\bibitem{paulraj2003introduction}
A.~Paulraj, R.~Nabar, D.~Gore, {Introduction to Space-Time Wireless
  Communications}, Cambridge University Press, 2003.

\bibitem{li2010multi}
H.~Li, A.~Attar, V.~Leung, {Multi-User Medium Access Control in Wireless Local
  Area Network}, in: Wireless Communications and Networking Conference (WCNC),
  2010 IEEE, IEEE, 2010, pp. 1--6.

\bibitem{cha2012performance}
J.~Cha, H.~Jin, B.~Jung, D.~Sung, Performance comparison of downlink user
  multiplexing schemes in ieee 802.11 ac: Multi-user mimo vs. frame
  aggregation, in: Wireless Communications and Networking Conference (WCNC),
  2012 IEEE, IEEE, 2012, pp. 1514--1519.

\bibitem{IEEE80211ac}
I.~P802.11ac/D4.0, {Draft Standard for Wireless LAN Medium Access Control (MAC)
  and Physical Layer (PHY) specifications Amendment 5: Enhancements for Very
  High Throughput for Operation in Bands below 6 GHz.}

\bibitem{akyildiz2007survey}
I.~Akyildiz, T.~Melodia, K.~Chowdhury, {A Survey on Wireless Multimedia Sensor
  Networks}, Computer Networks 51~(4) (2007) 921--960.

\bibitem{azadeh2011performance}
A.~Ettefagh, M.~Kuhn, C.~Eslin, A.~Wittneben, {Performance Analysis of
  Distributed Cluster-based MAC protocol for Multiuser MIMO Wireless Networks},
  EURASIP Journal on Wireless Communications and Networking 2011.

\bibitem{mecklenbrauker2011vehicular}
C.~Mecklenbrauker, A.~Molisch, J.~Karedal, F.~Tufvesson, A.~Paier, L.~Bernado,
  T.~Zemen, O.~Klemp, N.~Czink, {Vehicular Channel characterization and its
  Implications for Wireless system Design and Performance}, Proceedings of the
  IEEE 99~(7) (2011) 1189--1212.

\bibitem{IEEE80211b}
I.~802.11-2007, {Part 11: Wireless LAN Medium Access Control (MAC) and Physical
  Layer (PHY) Specifications: High-speed Physical Layer Extension in the 2.4
  GHz Band}.

\bibitem{gross1998fundamentals}
D.~Gross, C.~Harris, {Fundamentals of Queueing Systems}, John Wiley \& Sons,
  1998.

\bibitem{Bellalta2009-ASMTA}
B.~Bellalta, {A Queuing Model for the Non-continuous Frame Assembly Scheme in
  Finite Buffers}, in: Analytical and Stochastic Modeling Techniques and
  Applications, Lecture Notes in Computer Science, Springer, 2009, pp.
  219--233.

\bibitem{kuppa2006modeling}
S.~Kuppa, G.~Dattatreya, {Modeling and Analysis of Frame Aggregation in
  Unsaturated WLANs with Finite Buffer Stations}, in: IEEE International
  Conference on Communications, 2006, Vol.~3, IEEE, 2006, pp. 967--972.

\bibitem{lu2007performance}
K.~Lu, D.~Wu, Y.~Qian, Y.~Fang, R.~Qiu, {Performance of an Aggregation-based
  MAC Protocol for High-data-rate Ultra-wideband Ad-hoc Networks}, Vehicular
  Technology, IEEE Transactions on 56~(1) (2007) 312--321.

\bibitem{lu2009performance}
K.~Lu, J.~Wang, D.~Wu, Y.~Fang, {Performance of a Burst-frame-based CSMA/CA
  Protocol: Analysis and Enhancement}, Wireless Networks 15~(1) (2009) 87--98.

\bibitem{zhai2004performance}
H.~Zhai, Y.~Kwon, Y.~Fang, {Performance analysis of IEEE 802.11 MAC protocols
  in wireless LANs}, Wireless communications and mobile computing 4~(8) (2004)
  917--931.

\bibitem{lu2005performance}
K.~Lu, D.~Wu, Y.~Fang, R.~Qiu, {Performance Analysis of a Burst-frame-based MAC
  Protocol for Ultra-wideband A-hoc Networks}, in: Communications, 2005. ICC
  2005. 2005 IEEE International Conference on, Vol.~5, IEEE, 2005, pp.
  2937--2941.

\bibitem{bianchi2000performance}
G.~Bianchi, {Performance Analysis of the IEEE 802.11 Distributed Coordination
  Function}, Selected Areas in Communications, IEEE Journal on 18~(3) (2000)
  535--547.

\bibitem{tickoo2004queueing}
O.~Tickoo, B.~Sikdar, {Modeling Queueing and Channel Access Delay in
  Unsaturated IEEE 802.11 Random Access MAC based Wireless Networks}, IEEE/ACM
  Transactions on Networking (TON) 16~(4) (2008) 878--891.

\bibitem{grinstead1997introduction}
C.~Grinstead, J.~Snell, {Introduction to Probability}, American Mathematical
  Society, 1997.

\bibitem{chen2001component}
G.~Chen, B.~Szymanski, {Component-Oriented Simulation Architecture: Toward
  Interoperability and Interchangeability}, in: Simulation Conference, 2001.
  Proceedings of the Winter, Vol.~1, IEEE, 2001, pp. 495--501.

\bibitem{malone2007modeling}
D.~Malone, K.~Duffy, D.~Leith, {Modeling the 802.11 Distributed Coordination
  Function in Nonsaturated Heterogeneous Conditions}, IEEE/ACM Transactions on
  Networking 15~(1) (2007) 159--172.

\end{thebibliography}

\end{document}